\documentclass[a4paper]{article}\usepackage[]{graphicx}\usepackage[dvipsnames]{xcolor}
\makeatletter
\def\maxwidth{ %
  \ifdim\Gin@nat@width>\linewidth
    \linewidth
  \else
    \Gin@nat@width
  \fi
}
\makeatother

\definecolor{fgcolor}{rgb}{0.345, 0.345, 0.345}
\newcommand{\hlnum}[1]{\textcolor[rgb]{0.686,0.059,0.569}{#1}}%
\newcommand{\hlstr}[1]{\textcolor[rgb]{0.192,0.494,0.8}{#1}}%
\newcommand{\hlcom}[1]{\textcolor[rgb]{0.678,0.584,0.686}{\textit{#1}}}%
\newcommand{\hlopt}[1]{\textcolor[rgb]{0,0,0}{#1}}%
\newcommand{\hlstd}[1]{\textcolor[rgb]{0.345,0.345,0.345}{#1}}%
\newcommand{\hlkwa}[1]{\textcolor[rgb]{0.161,0.373,0.58}{\textbf{#1}}}%
\newcommand{\hlkwb}[1]{\textcolor[rgb]{0.69,0.353,0.396}{#1}}%
\newcommand{\hlkwc}[1]{\textcolor[rgb]{0.333,0.667,0.333}{#1}}%
\newcommand{\hlkwd}[1]{\textcolor[rgb]{0.737,0.353,0.396}{\textbf{#1}}}%

\usepackage{framed}
\makeatletter
\newenvironment{kframe}{%
 \def\at@end@of@kframe{}%
 \ifinner\ifhmode%
  \def\at@end@of@kframe{\end{minipage}}%
  \begin{minipage}{\columnwidth}%
 \fi\fi%
 \def\FrameCommand##1{\hskip\@totalleftmargin \hskip-\fboxsep
 \colorbox{shadecolor}{##1}\hskip-\fboxsep
     \hskip-\linewidth \hskip-\@totalleftmargin \hskip\columnwidth}%
 \MakeFramed {\advance\hsize-\width
   \@totalleftmargin\z@ \linewidth\hsize
   \@setminipage}}%
 {\par\unskip\endMakeFramed%
 \at@end@of@kframe}
\makeatother

\definecolor{shadecolor}{rgb}{.97, .97, .97}
\definecolor{messagecolor}{rgb}{0, 0, 0}
\definecolor{warningcolor}{rgb}{1, 0, 1}
\definecolor{errorcolor}{rgb}{1, 0, 0}
\newenvironment{knitrout}{}{} 

\usepackage{alltt}
\usepackage[a4paper, lmargin=3.5cm, rmargin=3cm]{geometry}
\usepackage{array}
\usepackage{times}
\usepackage[authoryear, round]{natbib}
\usepackage{booktabs}
\newcommand{\bcp}{{\bf P}}

\newcommand{\bct}{{\bf T}}

\newcommand{\bgamma}{\mbox{\boldmath $\Gamma$}}

\newcommand{\bfphi}{\mbox{\boldmath $\varphi$}}
\newcommand{\bfeta}{\mbox{\boldmath $\eta$}}

\newcommand{\btheta}{\mbox{\boldmath $\theta$}}
\newcommand{\balpha}{\mbox{\boldmath $\alpha$}}
\newcommand{\bfbeta}{\mbox{\boldmath $\beta$}}

\newcommand{\bfdelta}{\mbox{\boldmath $\delta$}}

\newcommand{\bone}{{\bf 1}}

\newcommand{\bflambda}{\mbox{\boldmath $\lambda$}}

\usepackage{graphicx}
\chardef\bslash=`\\ 

\usepackage{ulem} 

\newcommand{\hl}[1]{{#1}}

\usepackage{bm}
\usepackage{amsthm,amsmath,amssymb,amsfonts,amssymb}
\usepackage{listings}
\usepackage{enumerate}
\usepackage{enumitem}
\setlist[enumerate,1]{label={(\roman*)}}
\usepackage[figuresright]{rotating}
\usepackage{mathtools}
\usepackage[dvipsnames]{xcolor}
\usepackage{xurl}
\usepackage{hyperref}
\hypersetup{
colorlinks,
citecolor=black,
filecolor=black,
linkcolor=black,
urlcolor=black
}

\definecolor{backgroundColour}{rgb}{0.97,0.96,0.97}
\newcommand{\argmax}{\operatornamewithlimits{arg\,max}}

\IfFileExists{upquote.sty}{\usepackage{upquote}}{}
\begin{document}              
\makeatletter
\renewcommand{\maketitle}{\bgroup\setlength{\parindent}{0pt}
\begin{flushleft}
  {\large \textbf{\@title}}
  \\[4ex]
  \@author
\end{flushleft}\egroup

  \begingroup
    \renewcommand\thefootnote{\@fnsymbol\c@footnote}%
    \def\@makefnmark{\rlap{\@textsuperscript{\normalfont\@thefnmark}}}%
    \long\def\@makefntext##1{\parindent 1em\noindent
            \hb@xt@1.8em{%
                \hss\@textsuperscript{\normalfont\@thefnmark}}##1}%
    \thispagestyle{plain}\@thanks
  \endgroup
  \setcounter{footnote}{0}%
  \global\let\thanks\relax
  \global\let\maketitle\relax
  \global\let\@maketitle\relax
  \global\let\@thanks\@empty
  \global\let\@author\@empty
  \global\let\@date\@empty
  \global\let\@title\@empty
  \global\let\title\relax
  \global\let\author\relax
  \global\let\date\relax
  \global\let\and\relax
}
\makeatother

\renewcommand*{\thefootnote}{\fnsymbol{footnote}}

\title{A gentle tutorial on accelerated parameter and confidence interval estimation for hidden Markov models using Template Model Builder}

\date{}

\author{
  \noindent\textbf{Timoth\'ee Bacri} $^{1}$,
  \textbf{Geir D. Berentsen} $^{2}$,
  \textbf{Jan Bulla}\thanks{Corresponding author: e-mail: \sf{jan.bulla@uib.no}, phone: +47 55 58 28 75}\hspace{1ex}$^{1,3}$,
  \textbf{Sondre H{\o}lleland} $^{2,4}$\\[1ex]
  $^{1}$ \small{Department of Mathematics, University of Bergen, Postbox 7803, 5007 Bergen, Norway}\\
  $^{2}$ \small{Department of Business and Management Science, Norwegian School of Economics, Helleveien 30, 5045 Bergen, Norway}\\
  $^{3}$ \small{Department of Psychiatry and Psychotherapy, University of Regensburg, Universit\"atsstra{\ss}e 84, 93053 Regensburg, Germany}\\
  $^{4}$ \small{Department of Pelagic Fish, Institute of Marine Research, Postbox 1870, 5817 Bergen, Norway}
}

\maketitle


A very common way to estimate the parameters of a hidden Markov model (HMM) is the relatively straightforward computation of maximum likelihood (ML) estimates. For this task, most users rely on user-friendly implementation of the estimation routines via an interpreted programming language such as the statistical software environment {\tt{R}} \citep{rcoreteam}. Such an approach can easily require time-consuming computations, in particular for longer sequences of observations. In addition, selecting a suitable approach for deriving confidence intervals for the estimated parameters is not entirely obvious \citep[see, e.g.,][]{zucchini, lystig, visser}, and often the computationally intensive bootstrap methods have to be applied.

In this tutorial, we illustrate how to speed up the computation of ML estimates significantly via the {\tt{R}} package {\tt{TMB}}. Moreover, this approach permits simple retrieval of standard errors at the same time. We illustrate the performance of our routines using different data sets.
First, two smaller samples from a mobile application for tinnitus patients and a well-known data set of fetal lamb movements with 87 and 240 data points, respectively. Second, we rely on larger data sets of simulated data of sizes 2000 and 5000 for further analysis.
This tutorial is accompanied by a collection of scripts which are all available on GitHub. These scripts allow any user with moderate programming experience to benefit quickly from the computational advantages of {\tt{TMB}}.

\vspace*{1pc}
\noindent
\textit{Key words:} Hidden Markov model; {\tt TMB}; Confidence intervals; Maximum Likelihood Estimation; Tutorial\\
\\[2pt]
\noindent Supporting Information for this article is available from the authors or on the WWW under\break
\underline{\url{https://timothee-bacri.github.io/HMM_with_TMB}}



\newpage

\section{Introduction}
\label{sec:intro}

Hidden Markov models (HMMs) are a well-established, versatile type of model employed in many different applications. Since their first application in speech recognition \citep[see, e.g.,][]{baum, fredkin, gales}, HMMs found wide usage in many applied sciences. To name only a few, biology and bioinformatics \citep{schadt, durbin, eddy}, finance \citep{hamilton}, ecology \citep{mcclintock}, stochastic weather modeling \citep{lystig, ailliot}, and engineering \citep{mor}. Furthermore, the scientific literature on this topic in statistics is rich, as illustrated e.g.~by the manuscripts of \cite{zucchini, cappe, bartolucci}. The aforementioned sources contain, among many other aspects, detailed descriptions of parameter estimation for HMMs by maximization of the (log-)likelihood function. In short, maximum likelihood (ML) estimation is commonly achieved either by a direct numerical maximization as introduced by \citet{turner} and later detailed by \citet{zucchini}, who also provided a collection of {\tt{R}} \citep{rcoreteam} scripts that is widely used.
{\hl Alternatively, Expectation Maximization (EM) type algorithms as firstly described by \citet{bauma} or \citet{dempster} serve for parameter estimation equally well.
These algorithms possess several advantageous properties, for example, their robustness to poor initial values compared to direct numerical maximization via hill-climbing algorithms.
For more details on the EM algorithm in the context of HMMs and a comparison of both approaches, see \citet{bulla}, who also describe a hybrid approach combining both algorithms.}

Evaluating uncertainty and obtaining confidence intervals (CIs) constitutes another essential aspect when working with HMMs - and it is less straightforward than parameter estimation. Although \citet[][Ch.~12]{cappe} showed that CIs could be obtained based on asymptotic normality of the ML estimates of the parameters under certain conditions, \citet[p.~53]{fruhwirth-schnatter} points out that in independent mixture models, ``the regularity conditions are often violated". \citet[p.~68]{mclachlan} adds that ``In particular for mixture models, it is well known that the sample size $n$ has to be very large before the asymptotic theory of maximum likelihood applies." \citet{lystig} shows a way to compute the exact Hessian, and \citet{zucchini} presents an alternative way to compute the approximate Hessian and thus confidence intervals, but admits that ``the use of the Hessian to compute standard errors (and thence confidence intervals) is unreliable if some of the parameters are on or near the boundary of their parameter space". In addition, \citet{visser} report that computational problems arise when deriving CIs from the Hessian for sequences longer than 100 observations.

In this tutorial, we illustrate how to accelerate parameter estimation for HMMs with the help of Template Model Builder ({\tt{TMB}}). As described by \citet{kristensen}, {\tt{TMB}} is an {\tt{R}} package developed for efficiently fitting complex statistical models to data. It provides exact calculations of first- and second-order derivatives of the (log-)likelihood of a model by automatic differentiation, which allows for efficient gradient- and/or Hessian-based optimization of the likelihood on the one hand. On the other hand, {\tt{TMB}} permits to infer CIs for the estimated parameters by means of the Hessian. We show how to carry out this part for HMMs using a couple of simple examples. Then, we compare the Hessian-based CIs with CIs resulting from likelihood profiling and bootstrapping, which are both more computationally intensive.

The tutorial is accompanied by a collection of scripts {\hl (listed at \url{https://timothee-bacri.github.io/HMM_with_TMB/github.html\#directory-structure}\normalsize )}, which guide any user working with HMMs through the implementation of computationally efficient parameter estimation. The majority of scripts require only knowledge of {\tt{R}}, just the computation of the (log-)likelihood function requires the involvement of {\tt{C++}}. Moreover, we illustrate how {\tt{TMB}} permits Hessian- or profile likelihood-based CIs for the estimated parameters at a very low computational cost. Naturally, the accelerated parameter estimation procedure may also serve for implementing computationally more efficient bootstrap CIs, an aspect we also make use of for our analyses.

\newpage


\section{Principles of using {\tt{TMB}} for maximum likelihood estimation}
\label{sec:principles}

In order to keep this tutorial at acceptable length, all sections follow the same concept.
That is, the reader is encouraged to consult the respective part of our GitHub repository in parallel to reading each section; it is available at \url{https://timothee-bacri.github.io/HMM_with_TMB}. {\hl In particular, we recommend opening the file \textit{Data supplements.Rproj} (available in the related repository at \break \url{https://github.com/timothee-bacri/HMM_with_TMB}) with R-Studio, which lets the reader have the correct working path set up automatically.}
This permits to copy-paste or download all the scripts presented in this tutorial for each section on the one hand.
On the other hand, it allows for consistent maintenance of the code. Moreover, the repository also contains additional explanations, comments, and scripts.

\subsection{Setup}
\label{sec:setup}

Execution of our routines requires the installation of the {\tt{R}}-package {\tt{TMB}} and the software {\tt{Rtools}}, where the latter serves for compiling {\tt{C++}} code.
In order to ensure reproducibility of all results involving the generation of random numbers, the \texttt{set.seed} function requires {\tt{R}} version number 3.6.0 or greater.
Our scripts were tested on {\hl a workstation with 4 Intel(R) Xeon(R) Gold 6134 processors (3.7 GHz) each running under the Linux distribution Ubuntu 18.04.6 LTS (Bionic Beaver) with 384 GB RAM and required about one week of computing time.}

In particular for beginners, those parts of scripts involving {\tt{C++}} code can be difficult to debug because the code operates using a specific template.
Therefore it is helpful to know that {\tt{TMB}} provides a debugging feature, which can be useful to retrieve diagnostic error messages, in RStudio.
Enabling this feature is optional and can be achieved by the command \texttt{TMB:::setupRStudio()} (requires manual confirmation and re-starting RStudio).

\subsection{Linear regression example}
\label{sec:linreg}

We begin by demonstrating the principles of {\tt{TMB}}, which we illustrate through the fitting procedure for a simple linear model.
This permits, among other things, to show how to handle parameters subject to constraints, an aspect particularly relevant for HMMs.
A more comprehensive tutorial on {\tt{TMB}} presenting many technical details in more depths is available at \url{https://kaskr.github.io/adcomp/\_book/Tutorial.html}.

Let $\bm{x}$ and $\bm{y}$ denote the predictor and response vector, respectively, both of length $n$.
For a simple linear regression model with intercept $a$ and slope $b$, the negative log-likelihood equals
\begin{equation*}
- \log L(a, b, \sigma^2) = - \sum_{i=1}^n \log(\phi(y_i; a + bx_i, \sigma^2)),
\end{equation*}
where $\phi(\cdot; \mu, \sigma^2)$ corresponds to the density function of the univariate normal distribution with mean $\mu$ and variance $\sigma^2$.

The use of {\tt{TMB}} requires the (negative) log-likelihood function to be coded in {\tt{C++}} under a specific template, which is then loaded into {\tt{R}}.
The minimization of this function and other post-processing procedures are all carried out in {\tt{R}}.
Therefore, we require two files.
The first file, named \textit{linreg.cpp}, is written in {\tt{C++}} and defines the objective function, i.e.~the negative log-likelihood (nll) function of the linear model, as follows.\\

\begin{lstlisting}
#include <TMB.hpp> //import the TMB template

template<class Type>
Type objective_function<Type>::operator() ()
{
  DATA_VECTOR(y); // Data vector y passed from R
  DATA_VECTOR(x); // Data vector x passed from R
  
  PARAMETER(a);         // Parameter a passed from R 
  PARAMETER(b);         // Parameter b passed from R 
  PARAMETER(tsigma);    // Parameter sigma (transformed, on log-scale)
                        // passed from R
  
  // Transform tsigma back to natural scale
  Type sigma = exp(tsigma);
  
  // Declare negative log-likelihood
  Type nll = - sum(dnorm(y,
                         a + b * x,
                         sigma,
                         true));
  
  // Necessary for inference on sigma, not only tsigma
  ADREPORT(sigma);
  
  return nll;
}
\end{lstlisting}

Note that we define data inputs $x$ and $y$ using the \texttt{DATA\_VECTOR()} declaration in the above code.
Furthermore, we declare the nll as a function of the three parameters a, b and $\log(\sigma)$ using the \texttt{PARAMETER()} declaration.
In order to be able to carry out unconstrained optimization procedures in the following, the nll function is parametrized in terms of $\log(\sigma)$.
While the parameter $\sigma$ is constrained to be non-negative, $\log(\sigma)$ can be freely estimated.
Alternatively, constraint optimization methods could be carried out, but we do not investigate such procedures.
The \texttt{ADREPORT()} function is optional but useful for parameter inference at the postprocessing stage.

The second file needed is written in {\tt{R}} and serves for compiling the nll function defined above and carrying out the estimation procedure by numerical optimization of the nll function.
The .R file (shown below) carries out the compilation of the {\tt{C++}} file and minimization of the nll function:
\begin{knitrout}
\definecolor{shadecolor}{rgb}{0.969, 0.969, 0.969}\color{fgcolor}\begin{kframe}
\begin{alltt}
\hlcom{# Loading TMB package}
\hlkwd{library}\hlstd{(TMB)}
\hlcom{# Compilation. The compiler returns 0 if the compilation of}
\hlcom{# the cpp file was successful}
\hlstd{TMB}\hlopt{::}\hlkwd{compile}\hlstd{(}\hlstr{"code/linreg.cpp"}\hlstd{)}
\end{alltt}
\begin{verbatim}
## [1] 0
\end{verbatim}
\begin{alltt}
\hlcom{# Dynamic loading of the compiled cpp file}
\hlkwd{dyn.load}\hlstd{(}\hlkwd{dynlib}\hlstd{(}\hlstr{"code/linreg"}\hlstd{))}
\hlcom{# Generate the data for our test sample}
\hlkwd{set.seed}\hlstd{(}\hlnum{123}\hlstd{)}
\hlstd{data} \hlkwb{<-} \hlkwd{list}\hlstd{(}\hlkwc{y} \hlstd{=} \hlkwd{rnorm}\hlstd{(}\hlnum{20}\hlstd{)} \hlopt{+} \hlnum{1}\hlopt{:}\hlnum{20}\hlstd{,} \hlkwc{x} \hlstd{=} \hlnum{1}\hlopt{:}\hlnum{20}\hlstd{)}
\hlstd{parameters} \hlkwb{<-} \hlkwd{list}\hlstd{(}\hlkwc{a} \hlstd{=} \hlnum{0}\hlstd{,} \hlkwc{b} \hlstd{=} \hlnum{0}\hlstd{,} \hlkwc{tsigma} \hlstd{=} \hlnum{0}\hlstd{)}
\hlcom{# Instruct TMB to create the likelihood function}
\hlstd{obj_linreg} \hlkwb{<-} \hlkwd{MakeADFun}\hlstd{(data, parameters,} \hlkwc{DLL} \hlstd{=} \hlstr{"linreg"}\hlstd{,}
                        \hlkwc{silent} \hlstd{=} \hlnum{TRUE}\hlstd{)}
\hlcom{# Optimization of the objective function with nlminb}
\hlcom{# Arguments are the initial parameters, the objective function to}
\hlcom{# minimize, and the gradient and Hessian passed from TMB}
\hlstd{mod_linreg} \hlkwb{<-} \hlkwd{nlminb}\hlstd{(}\hlkwc{start} \hlstd{= obj_linreg}\hlopt{$}\hlstd{par,}
                     \hlkwc{objective} \hlstd{= obj_linreg}\hlopt{$}\hlstd{fn,}
                     \hlkwc{gradient} \hlstd{= obj_linreg}\hlopt{$}\hlstd{gr,}
                     \hlkwc{hessian} \hlstd{= obj_linreg}\hlopt{$}\hlstd{he)}
\hlstd{mod_linreg}\hlopt{$}\hlstd{par}
\end{alltt}
\begin{verbatim}
##           a           b      tsigma 
##  0.31009251  0.98395536 -0.05814649
\end{verbatim}
\end{kframe}
\end{knitrout}

In addition to the core functionality presented above, different types of post-processing of the results are possible as well. For example, the function \texttt{sdreport} returns the ML estimates and standard errors of the parameters in terms of which the nll is parametrized:
\begin{knitrout}
\definecolor{shadecolor}{rgb}{0.969, 0.969, 0.969}\color{fgcolor}\begin{kframe}
\begin{alltt}
\hlkwd{sdreport}\hlstd{(obj_linreg,} \hlkwc{par.fixed} \hlstd{= mod_linreg}\hlopt{$}\hlstd{par)}
\end{alltt}
\begin{verbatim}
## sdreport(.) result
##           Estimate Std. Error
## a       0.31009251 0.43829087
## b       0.98395536 0.03658782
## tsigma -0.05814649 0.15811383
## Maximum gradient component: 1.300261e-10
\end{verbatim}
\end{kframe}
\end{knitrout}

In principle, the argument \texttt{par.fixed = mod\_linreg\$par} is optional but recommended, because it ensures that the \texttt{sdreport} function is carried out at the minimum found by {\tt{nlminb}} \citep{gay}.
{\hl Instead of \texttt{nlminb}, other optimization routines may be used as well.
For example \citet{zucchini} rely on the {\tt{R}} function \texttt{nlm}.
We selected \texttt{nlminb} because it offers a higher degree of flexibility, while having a syntax close to \texttt{nlm}.}
Note that the standard errors above are based on the Hessian matrix of the nll.

From a practical perspective, it is usually desirable to obtain standard errors for the constrained variables, in this case $\sigma$. To achieve this, one can run the \texttt{summary} function with argument \texttt{select = "report"}:
\begin{knitrout}
\definecolor{shadecolor}{rgb}{0.969, 0.969, 0.969}\color{fgcolor}\begin{kframe}
\begin{alltt}
\hlkwd{summary}\hlstd{(}\hlkwd{sdreport}\hlstd{(obj_linreg,} \hlkwc{par.fixed} \hlstd{= mod_linreg}\hlopt{$}\hlstd{par),}
        \hlkwc{select} \hlstd{=} \hlstr{"report"}\hlstd{)}
\end{alltt}
\begin{verbatim}
##        Estimate Std. Error
## sigma 0.9435117  0.1491823
\end{verbatim}
\end{kframe}
\end{knitrout}
These standard errors result from the generalized delta method described by \citet{kass}, which is implemented within {\tt{TMB}}. Note that full functionality of the \texttt{sdreport} function requires calling the function \texttt{ADREPORT} on the additional parameters of interest (i.e. those including transformed parameters, in our example $\sigma$) in the {\tt{C++}} part.
The \texttt{select} argument restricts the output to variables passed by \texttt{ADREPORT}.
This feature is particularly useful when the likelihood has been reparametrized as above, and is especially relevant for HMMs.
Following \citet{zucchini}, we refer to the original parameters as natural parameters, and to their transformed version as the working parameters.

Last, we display the estimation results from the \texttt{lm} function for comparison.
\begin{knitrout}
\definecolor{shadecolor}{rgb}{0.969, 0.969, 0.969}\color{fgcolor}\begin{kframe}
\begin{alltt}
\hlkwd{rbind}\hlstd{(}
  \hlstr{"lm"}  \hlstd{=} \hlkwd{lm}\hlstd{(y} \hlopt{~} \hlstd{x,} \hlkwc{data} \hlstd{= data)}\hlopt{$}\hlstd{coef,} \hlcom{# linear regression using R}
  \hlstr{"TMB"} \hlstd{= mod_linreg}\hlopt{$}\hlstd{par[}\hlnum{1}\hlopt{:}\hlnum{2}\hlstd{]} \hlcom{# intercept and slope from TMB fit}
\hlstd{)}
\end{alltt}
\begin{verbatim}
##     (Intercept)         x
## lm    0.3100925 0.9839554
## TMB   0.3100925 0.9839554
\end{verbatim}
\end{kframe}
\end{knitrout}

\section{Parameter estimation techniques for HMMs}
\label{sec:estimation}

In this section we recall basic concepts underlying parameter estimation for HMMs via direct numerical optimization of the likelihood. In terms of notation, we stay as close as possible to \citet{zucchini}, where a more detailed presentation is available.

\subsection{Basic notation and model setup}
\label{sec:notation}

\noindent A large variety of modeling approaches is possible with HMMs, ranging from rather simple to highly complex setups. In a basic HMM, one assumes that the data-generating process corresponds to a time-dependent mixture of conditional distributions. More specifically, the mixing process is driven by an unobserved (hidden) homogeneous Markov chain. In this paper we focus on a Poisson HMM, but only small changes are necessary to adapt our scripts to models with other conditional distributions. {\hl For an example of the application of {\tt{TMB}} in a more complex HMM setting, see \cite{berentsen}.}  Let $\{X_t: t = 1, \ldots, T\}$ and $\{C_t : t = 1, \ldots, T\}$ denote the observed and hidden process, respectively, where $t$ denotes the (time) index ranging from one to $T$.
For an $m$-state Poisson HMM, the conditional distributions with parameter $\lambda_i$ are then specified through
\begin{equation*}
p_i(x) = \text{P}(X_t = x \vert C_t = i) = \frac{e^{-\lambda_i} \lambda_i^x}{x!},
\end{equation*}
where $i = 1, \ldots, m$. Furthermore, we let $\bgamma = \{\gamma_{ij}\}$ and $\bfdelta$ denote the transition probability matrix (TPM) of the Markov chain and the corresponding stationary distribution, respectively.
It is noteworthy that Markov chains in the context of HMMs are often assumed irreducible and aperiodic.
For example,
\citet[Lemma 6.3.5 on p.~225 and Theorem 6.4.3 on p.~227]{grimmett} show that irreducibility ensures the existence of the stationary distribution, and \citet[p.~394]{feller} describe that aperiodicity implies that a unique limiting distribution exists and corresponds to the stationary distribution.
These results are, however, of limited relevance for most estimation algorithms, because the elements of $\bgamma$ are in general strictly positive. Nevertheless, one should be careful when manually fixing selected elements of $\bgamma$ to zero.

\subsection{The likelihood function of an HMM}
\label{sec:hmm_likelihood}

The likelihood function of an HMM requires, in principle, a summation over all possible state sequences. As shown e.g.~by \citet[p.~37]{zucchini}, a computationally convenient representation as a product of matrices is possible. Let $X^{(t)} = \{X_1, \ldots, X_t \}$ and $x^{(t)} = \{x_1, \ldots, x_t \}$ denote the 'history' of the observed process $X_t$ and the observations $x_t$, respectively, from time one up to time $t$. Moreover, let $\btheta$ denote the vector of model parameters, which consists of the parameters of the TPM and the parameters of the conditional probability density functions. Given these parameters, the likelihood of the observations $\{x_1, \ldots, x_T \}$ can then be expressed as
\begin{equation}
\label{eq:hmm_likelihood}
L(\btheta) = \bcp(X^{(T)} = x^{(T)}) = \bfdelta \bcp(x_1) \bgamma \bcp(x_2) \bgamma \bcp(x_3) \ldots \bgamma \bcp(x_T) \bone',
\end{equation}
where
\begin{equation*}
\bcp(x) = \begin{pmatrix}
p_1(x)    &         &         & 0\\
          & p_2(x)  &         &\\
          &         & \ddots  &\\
0         &         &         & p_m(x)
\end{pmatrix}
\end{equation*}
corresponds to a diagonal matrix with the $m$ conditional probability density functions evaluated at $x$ (we will use the term density despite the discrete support), and $\bone$ denotes a vector of ones. The first element of the likelihood function, the so-called initial distribution, is given by the stationary distribution $\bfdelta$ here. Alternatively, the initial distribution may be estimated freely, which requires minor changes to the likelihood function discussed in \autoref{sec:tmb_cpp}.

Note that the treatment of missing data is comparably straightforward in this setup. If $x$ is a missing observation, one just has to set $p_i(x) =  1$, thus $\bcp(x)$ reduces to the unity matrix as detailed in \citet[p.~40]{zucchini}.
\citet[p.~41]{zucchini} also explains how to adjust the likelihood when entire intervals are missing. Furthermore, this representation of the likelihood is quite natural from an intuitive point of view. From left to right, it can be interpreted as a pass through the observations: one starts with the initial distribution multiplied by the conditional density of $x_1$ collected in $\bcp(x_1)$. This is followed by iterative multiplications with the TPM modeling the transition to the next observation, and yet another multiplication with contributions of the following conditional densities.

\subsection{Forward algorithm and backward algorithm}
\label{sec:hmm_fwbw}

The pass through the observations described above actually forms the basis for an efficient evaluation of the likelihood function. More precisely, the so-called ``forward algorithm" allows for a recursive computation of the likelihood. For setting up this algorithm, we need to define the vector $\balpha_t$ by
\begin{align*}
\balpha_t &= \bfdelta \bcp(x_1)\bgamma \bcp(x_2) \bgamma \bcp(x_3) \ldots \bgamma \bcp(x_t)\\
&= \bfdelta \bcp(x_1) \prod_{s=2}^{t}\bgamma \bcp(x_s)\\
&= \left( \alpha_t(1), \ldots, \alpha_t(m) \right)
\end{align*}
for $t = 1, 2, \ldots, T$. 
The name forward algorithm originates from the way of calculating $\balpha_t$, i.e.
\begin{gather*}
\balpha_0 = \bfdelta \bcp(x_1)\\
\balpha_t = \balpha_{t-1} \bgamma \bcp(x_t) \text{ for } t = 1, 2, \ldots, T.
\end{gather*}
After a pass through all observations, the likelihood results from
\begin{gather*}
L(\btheta) = \balpha_T \bone'.
\end{gather*}
In a similar way, the ``backward algorithm" also permits the recursive computation of the likelihood, but starting with the last observation. To formulate the backward algorithm, let us define the vector $\bfbeta_t$ for $t = 1, 2,
\ldots, T$ so that
\begin{align*}
\bfbeta'_t &= \bgamma \bcp(x_{t+1}) \bgamma \bcp(x_{t+2}) \ldots \bgamma \bcp(x_T) \ldots \bone'\\
&= \left(\prod_{s=t+1}^{T}\bgamma \bcp(x_s) \right) \bone'\\
&= \left( \beta_t(1), \ldots, \beta_t(m) \right).
\end{align*}
The name backward algorithm results from the way of calculating $\bfbeta_t$, i.e.
\begin{gather*}
\bfbeta_T = \bone'\\
\bfbeta_t = \bgamma \bcp(x_{t+1}) \bfbeta_{t+1} \text{ for } t = T-1, T-2, \ldots, 1.
\end{gather*}
Again, the likelihood can be calculated after a pass through all observations by
\begin{gather*}
L(\btheta) = \bfdelta \bfbeta_1.
\end{gather*}
In general, parameter estimation bases on the forward algorithm.
The backward algorithm is, however, still useful because the quantities $\balpha_t$ and $\bfbeta_t$ together serve for a couple of interesting tasks.
For example, they are the basis for deriving a particular type of conditional distributions and for state inference by local decoding \cite[Ch.~5, pp.~81-93]{zucchini}. We present details on local decoding on the GitHub page.

Last, it is well-known that the execution of the forward (or backward) algorithm may quickly lead to underflow errors, because many elements of the vectors and matrices involved take values between zero and one. To avoid these difficulties, a scaling factor can be introduced. We follow the approach suggested by \citet[p.~48]{zucchini} and implement a scaled version of the forward algorithm, which directly provides the (negative) log-likelihood as result.

\subsection{Reparametrization of the likelihood function}
\label{sec:hmm_repar}

The representation of the likelihood and the algorithms shown above rely on the data and the set of parameters $\btheta$ as input. The data are subject to several constraints:
\begin{enumerate}
\item Typically there are various constraints of the parameters in the conditional distribution. For the Poisson HMM, all elements of the parameter vector $\bflambda\ = (\lambda_1, \dots, \lambda_m)$ must be non-negative. 
\item In general, the parameters $\gamma_{ij}$ of the TPM $\bgamma$ have to be non-negative, and the rows of $\bgamma$ must sum up to one.
\end{enumerate}
The constraints of the TPM can be difficult to deal with using constrained optimization of the likelihood. A common approach is to reparametrize the log-likelihood in terms of unconstrained ``working" parameters $\{\bct, \bfeta\}= g^{-1}(\bgamma, \bflambda)$, as follows. A possible reparametrization of $\bgamma$ is given by 

\begin{equation*}
\gamma_{ij} = \frac{\exp(\tau_{ij})}{1 + \sum_{k \neq i} \exp(\tau_{ik})}, \text{ for } i \neq j,
\end{equation*}
where $\tau_{ij}$ are $m(m-1)$ real-valued, thus unconstrained, elements of an $m$ times $m$ matrix $\bct$ with no diagonal elements. The diagonal elements of $\bgamma$ follows implicitly from $\sum_j \gamma_{ij} = 1 \;\forall\; i$ \cite[p.~51]{zucchini}. The corresponding reverse transformation is given by
\begin{equation*}
\tau_{ij} = \log\left(\frac{\gamma_{ij}}{1 - \sum_{k \neq i} \gamma_{ik}}\right) = \log(\gamma_{ij}/\gamma_{ii}), \text{ for } i \neq j.
\end{equation*}

For the Poisson HMM the intensities can be reparametrized in terms of $\lambda_i = \exp(\eta_i)$, and consequently the unconstrained working parameters are given by $\eta_i = \log(\lambda_i), i = 1,\dots,m$. Estimates of the ``natural" parameters $\{\bgamma, \bflambda\}$ can then be obtained by maximizing the reparametrized likelihood with respect to $\{\bct, \bfeta\}$ and then transforming the estimated working parameters back to natural parameters via the above transformations, i.e. $\{\hat{\bgamma}, \hat{\bflambda}\} = g(\hat{\bct}, \hat{\bfeta})$. Note that in general the function $g$ needs to be one-to-one for the above procedure to work.

\section{Using TMB}
\label{sec:tmb}

In the following we show how ML estimation of the parameters of HMMs can be carried out efficiently via {\tt{TMB}}. 

\subsection{Likelihood function}
\label{sec:tmb_cpp}

Similar to the linear regression example presented in \autoref{sec:linreg}, the first and essential step is to define our nll function to be minimized later in a suitable {\tt{C++}} file. In our case, this function calculates the negative log-likelihood presented by \citet[p.~48]{zucchini}, and our {\tt{C++}} code is analog to the {\tt{R}}-code shown by \citet[p.~331 - 333]{zucchini}. This function, defined in the file named \textit{poi\_hmm.cpp}, tackles our setting with conditional Poisson distributions only. An extension to, e.g., Gaussian, binomial and exponential conditional distributions is straightforward. It requires to modify the density function in the \textit{poi\_hmm.cpp} file and the related functions for parameter transformation presented in \autoref{sec:hmm_repar}. We illustrate the implementation of the Gaussian case in the GitHub repository.  

However, note that the number of possible modelling setups is very large: e.g., the conditional distributions may vary from state to state, nested model specifications, the conditional mean may be linked to covariates, or the TPM could depend on covariates - to name only a few. Due to the very large number of possible extensions of the basic HMM, we refrain from implementing an {\tt{R}}-package, but prefer to provide a proper guidance to the reader for building custom models suited to a particular application. As a small example, we illustrate how to implement a freely estimated initial distribution in the file \textit{poi\_hmm.cpp}. This modification can be achieved by uncommenting a couple of lines only.

{\hl
The file \textit{poi\_hmm.cpp} is available at {\url{https://github.com/timothee-bacri/HMM_with_TMB/blob/main/code/poi_hmm.cpp} and contains the following.}
}
\begin{lstlisting}
#include <TMB.hpp>
#include "../functions/utils.cpp"

// Likelihood for a poisson hidden markov model. 
template<class Type>
Type objective_function<Type>::operator() ()
{
  
  // Data
  DATA_VECTOR(x);          // timeseries vector
  DATA_INTEGER(m);         // Number of states m
  
  // Parameters
  PARAMETER_VECTOR(tlambda);     // conditional log_lambdas's
  PARAMETER_VECTOR(tgamma);      // m(m-1) working parameters of TPM
  
  // Uncomment only when using a non-stationary distribution
  //PARAMETER_VECTOR(tdelta);    // transformed stationary distribution,
  
  // Transform working parameters to natural parameters:
  vector<Type> lambda = tlambda.exp();
  matrix<Type> gamma = gamma_w2n(m, tgamma);
  
  // Construct stationary distribution
  vector<Type> delta = stat_dist(m, gamma);
  // If using a non-stationary distribution, use this instead
  //vector<Type> delta = delta_w2n(m, tdelta);
  
  // Get number of timesteps (n)
  int n = x.size();
  
  // Evaluate conditional distribution: Put conditional
  // probabilities of observed x in n times m matrix
  // (one column for each state, one row for each datapoint):
  matrix<Type> emission_probs(n, m);
  matrix<Type> row1vec(1, m);
  row1vec.setOnes();
  for (int i = 0; i < n; i++) {
    if (x[i] != x[i]) { // f != f returns true if and only if f is NaN. 
      // Replace missing values (NA in R, NaN in C++) with 1
      emission_probs.row(i) = row1vec;
    }
    else {
      emission_probs.row(i) = dpois(x[i], lambda, false);
    }
  }
  
  // Corresponds to (Zucchini et al., 2016, p 333)
  matrix<Type> foo, P;
  Type mllk, sumfoo, lscale;
  
  foo = (delta * vector<Type>(emission_probs.row(0))).matrix();
  sumfoo = foo.sum();
  lscale = log(sumfoo);
  foo.transposeInPlace();
  foo /= sumfoo;
  for (int i = 2; i <= n; i++) {
    P = emission_probs.row(i - 1);
    foo = ((foo * gamma).array() * P.array()).matrix();
    sumfoo = foo.sum();
    lscale += log(sumfoo);
    foo /= sumfoo;
  }
  mllk = -lscale;
  
  // Use adreport on variables for which we want standard errors
  ADREPORT(lambda);
  ADREPORT(gamma);
  ADREPORT(delta);
  
  // Variables we need for local decoding and in a convenient format
  REPORT(lambda);
  REPORT(gamma);
  REPORT(delta);
  REPORT(n);
  REPORT(emission_probs);
  REPORT(mllk);
  
  return mllk;
}
\end{lstlisting}

\subsection{Optimization}
\label{sec:tmb_r}

With the nll function available in {\tt{C++}}, we can carry out the parameter estimation and all pre-/post-processing in {\tt{R}}. In the following we describe the steps to be carried out.

\begin{enumerate}
\item Loading of the necessary packages, compilation of the nll function with {\tt{TMB}} and subsequent loading, and loading of the auxiliary functions for parameter transformation.
\begin{knitrout}
\definecolor{shadecolor}{rgb}{0.969, 0.969, 0.969}\color{fgcolor}\begin{kframe}
\begin{alltt}
\hlcom{# Load TMB and optimization packages}
\hlkwd{library}\hlstd{(TMB)}
\hlcom{# Run the C++ file containing the TMB code}
\hlstd{TMB}\hlopt{::}\hlkwd{compile}\hlstd{(}\hlstr{"code/poi_hmm.cpp"}\hlstd{)}
\end{alltt}
\begin{verbatim}
## [1] 0
\end{verbatim}
\begin{alltt}
\hlcom{# Load it}
\hlkwd{dyn.load}\hlstd{(}\hlkwd{dynlib}\hlstd{(}\hlstr{"code/poi_hmm"}\hlstd{))}
\hlcom{# Load the parameter transformation function}
\hlkwd{source}\hlstd{(}\hlstr{"functions/utils.R"}\hlstd{)}
\end{alltt}
\end{kframe}
\end{knitrout}

\item Loading of the observations. The data are part of a large data set collected with the ``Track Your Tinnitus" (TYT) mobile application, a detailed description of which is presented in \citet{pryss} and \citet{pryssa}.
We analyze 87 successive days of the ``arousal" variable recorded for a single individual. This variable is measured on a discrete scale, where higher values correspond to a higher degree of excitement and lower values to a more calm emotional state \citep[for details, see][]{probst, probsta}.

Loading the ``arousal" variable can be achieved simply with
\begin{knitrout}
\definecolor{shadecolor}{rgb}{0.969, 0.969, 0.969}\color{fgcolor}\begin{kframe}
\begin{alltt}
\hlkwd{load}\hlstd{(}\hlstr{"data/tinnitus.RData"}\hlstd{)}
\end{alltt}
\end{kframe}
\end{knitrout}

\autoref{table:tinnitus_data} presents the raw data, which are also available for download at the GitHub repository.

\begin{table}[ht]
\centering
\begin{tabular}{p{15cm}}
   \hline
6 5 3 6 4 3 5 6 6 6 4 6 6 4 6 6 6 6 6 4 6 5 6 7 6 5 5 5 7 6 5 6 5 6 6 6 5 6 7 7 6 7 6 6 6 6 5 7 6 1 6 0 2 1 6 7 6 6 6 5 5 6 6 2 5 0 1 1 1 2 3 1 3 1 3 0 1 1 1 4 1 4 1 2 2 2 0 \\ 
   \hline
\end{tabular}
\caption{TYT data. Observations collected by the TYT app on 87 successive days (from left to right) for a single individual.} 
\label{table:tinnitus_data}
\end{table}


\item Initialization of the number of states and starting (or initial) values for the optimization. First, the number of states needs to be determined. As explained by \citet{pohlea}, \citet{pohle}, and \citet[][Section 6]{zucchini} (to name only a few), usually one would first fit models with a different number of states. Then, these models are evaluated e.g.~by means of model selection criteria \citep[as carried out by][]{leroux} or prediction performance \citep{celeux}. The model selection procedure shows that both AIC and BIC prefer a two-state model over a model with three or four states. Consequently, we focus on the two-state model in the following.

The list object \texttt{TMB\_data} contains the data and the number of states.
\begin{knitrout}
\definecolor{shadecolor}{rgb}{0.969, 0.969, 0.969}\color{fgcolor}\begin{kframe}
\begin{alltt}
\hlcom{# Model with 2 states}
\hlstd{m} \hlkwb{<-} \hlnum{2}
\hlstd{TMB_data} \hlkwb{<-} \hlkwd{list}\hlstd{(}\hlkwc{x} \hlstd{= tinn_data,} \hlkwc{m} \hlstd{= m)}
\end{alltt}
\end{kframe}
\end{knitrout}
Secondly, initial values for the optimization procedure need to be defined. Although we will apply unconstrained optimization, we initialize the natural parameters, because this is much more intuitive and practical than handling the working parameters. 
\begin{knitrout}
\definecolor{shadecolor}{rgb}{0.969, 0.969, 0.969}\color{fgcolor}\begin{kframe}
\begin{alltt}
\hlcom{# Generate initial set of parameters for optimization}
\hlstd{lambda} \hlkwb{<-} \hlkwd{c}\hlstd{(}\hlnum{1}\hlstd{,} \hlnum{3}\hlstd{)}
\hlstd{gamma} \hlkwb{<-} \hlkwd{matrix}\hlstd{(}\hlkwd{c}\hlstd{(}\hlnum{0.8}\hlstd{,} \hlnum{0.2}\hlstd{,}
                  \hlnum{0.2}\hlstd{,} \hlnum{0.8}\hlstd{),} \hlkwc{byrow} \hlstd{=} \hlnum{TRUE}\hlstd{,} \hlkwc{nrow} \hlstd{= m)}
\end{alltt}
\end{kframe}
\end{knitrout}

\item Transformation from natural to working parameters. The previously created initial values are transformed and stored in the list {\tt parameters} for the optimization procedure.
\begin{knitrout}
\definecolor{shadecolor}{rgb}{0.969, 0.969, 0.969}\color{fgcolor}\begin{kframe}
\begin{alltt}
\hlcom{# Turn them into working parameters}
\hlstd{parameters} \hlkwb{<-} \hlkwd{pois.HMM.pn2pw}\hlstd{(m, lambda, gamma)}
\end{alltt}
\end{kframe}
\end{knitrout}

\item Creation of the \texttt{TMB} negative log-likelihood function with its derivatives. This object, stored as \texttt{obj\_tmb} requires the data, the initial values, and the previously created DLL as input. Setting argument \texttt{silent = TRUE} disables tracing information and is only used here to avoid excessive output.

\begin{knitrout}
\definecolor{shadecolor}{rgb}{0.969, 0.969, 0.969}\color{fgcolor}\begin{kframe}
\begin{alltt}
\hlstd{obj_tmb} \hlkwb{<-} \hlkwd{MakeADFun}\hlstd{(TMB_data, parameters,}
                     \hlkwc{DLL} \hlstd{=} \hlstr{"poi_hmm"}\hlstd{,} \hlkwc{silent} \hlstd{=} \hlnum{TRUE}\hlstd{)}
\end{alltt}
\end{kframe}
\end{knitrout}

This object also contains the previously defined initial values as a vector (\texttt{par}) rather than a list. The negative log-likelihood (\texttt{fn}), its gradient (\texttt{gr}), and Hessian (\texttt{he}) are functions of the parameters (in vector form) while the data are considered fixed: 

\begin{knitrout}
\definecolor{shadecolor}{rgb}{0.969, 0.969, 0.969}\color{fgcolor}\begin{kframe}
\begin{alltt}
\hlstd{obj_tmb}\hlopt{$}\hlstd{par}
\end{alltt}
\begin{verbatim}
##   tlambda   tlambda    tgamma    tgamma 
##  0.000000  1.098612 -1.386294 -1.386294
\end{verbatim}
\begin{alltt}
\hlstd{obj_tmb}\hlopt{$}\hlkwd{fn}\hlstd{(obj_tmb}\hlopt{$}\hlstd{par)}
\end{alltt}
\begin{verbatim}
## [1] 228.3552
\end{verbatim}
\begin{alltt}
\hlstd{obj_tmb}\hlopt{$}\hlkwd{gr}\hlstd{(obj_tmb}\hlopt{$}\hlstd{par)}
\end{alltt}
\begin{verbatim}
##          [,1]      [,2]     [,3]      [,4]
## [1,] -3.60306 -146.0336 10.52832 -1.031706
\end{verbatim}
\begin{alltt}
\hlstd{obj_tmb}\hlopt{$}\hlkwd{he}\hlstd{(obj_tmb}\hlopt{$}\hlstd{par)}
\end{alltt}
\begin{verbatim}
##           [,1]       [,2]       [,3]       [,4]
## [1,]  1.902009  -5.877900 -1.3799682  2.4054017
## [2,] -5.877900 188.088247 -4.8501589  2.3434284
## [3,] -1.379968  -4.850159  9.6066700 -0.8410438
## [4,]  2.405402   2.343428 -0.8410438  0.7984216
\end{verbatim}
\end{kframe}
\end{knitrout}

\item Execution of the optimization. For this step we rely again on the optimizer implemented in the {\tt{nlminb}} function. The arguments, i.e.~ initial values for the parameters and the function to be optimized, are extracted from the previously created {\tt{TMB}} object. 
\begin{knitrout}
\definecolor{shadecolor}{rgb}{0.969, 0.969, 0.969}\color{fgcolor}\begin{kframe}
\begin{alltt}
\hlstd{mod_tmb} \hlkwb{<-} \hlkwd{nlminb}\hlstd{(}\hlkwc{start} \hlstd{= obj_tmb}\hlopt{$}\hlstd{par,} \hlkwc{objective} \hlstd{= obj_tmb}\hlopt{$}\hlstd{fn)}
\hlcom{# Check that it converged successfully}
\hlstd{mod_tmb}\hlopt{$}\hlstd{convergence} \hlopt{==} \hlnum{0}
\end{alltt}
\begin{verbatim}
## [1] TRUE
\end{verbatim}
\end{kframe}
\end{knitrout}
As mentioned previously, various alternatives to {\tt{nlminb}} exist and could be used at this step instead.

\item Obtaining the ML estimates of the natural parameters together with their standard errors is possible by using the previously introduced command \texttt{sdreport}. Recall that this requires the parameters of interest to be treated by the \texttt{ADREPORT} statement in the {\tt{C++}} part. It should be noted that the presentation of the set of parameters \texttt{gamma} below results from a column-wise representation of the TPM.
\begin{knitrout}
\definecolor{shadecolor}{rgb}{0.969, 0.969, 0.969}\color{fgcolor}\begin{kframe}
\begin{alltt}
\hlkwd{summary}\hlstd{(}\hlkwd{sdreport}\hlstd{(obj_tmb,} \hlkwc{par.fixed} \hlstd{= mod_tmb}\hlopt{$}\hlstd{par),} \hlstr{"report"}\hlstd{)}
\end{alltt}
\begin{verbatim}
##          Estimate Std. Error
## lambda 1.63641070 0.27758294
## lambda 5.53309626 0.31876141
## gamma  0.94980192 0.04374682
## gamma  0.02592209 0.02088689
## gamma  0.05019808 0.04374682
## gamma  0.97407791 0.02088689
## delta  0.34054163 0.23056401
## delta  0.65945837 0.23056401
\end{verbatim}
\end{kframe}
\end{knitrout}
Note that the table above also contains estimation results for $\bfdelta$ and accompanying standard errors, although $\bfdelta$ is not estimated, but derived from $\bgamma$. We provide further details on this aspect in \autoref{sec:hessian}.

The value of the nll function in the minimum found by the optimizer can also be extracted directly from the object {\tt mod\_tmb} by accessing the list element {\tt objective}:
\begin{knitrout}
\definecolor{shadecolor}{rgb}{0.969, 0.969, 0.969}\color{fgcolor}\begin{kframe}
\begin{alltt}
\hlstd{mod_tmb}\hlopt{$}\hlstd{objective}
\end{alltt}
\begin{verbatim}
## [1] 168.5361
\end{verbatim}
\end{kframe}
\end{knitrout}

\item In the optimization above we already benefited from an increased speed due to the evaluation of the nll in {\tt{C++}} compared to the forward algorithm being executed entirely in {\tt{R}}. However, the use of {\tt{TMB}} also permits to introduce the gradient and/or the Hessian computed by {\tt{TMB}} into the optimization procedure. This is in general advisable, because {\tt{TMB}} provides an exact value of both gradient and Hessian up to machine precision, which is superior to approximations used by optimizing procedure. Similar to the nll, both quantities can be extracted directly from the {\tt{TMB}} object {\tt obj\_tmb}:
\begin{knitrout}
\definecolor{shadecolor}{rgb}{0.969, 0.969, 0.969}\color{fgcolor}\begin{kframe}
\begin{alltt}
\hlcom{# The negative log-likelihood is accessed by the objective}
\hlcom{# attribute of the optimized object}
\hlstd{mod_tmb} \hlkwb{<-} \hlkwd{nlminb}\hlstd{(}\hlkwc{start} \hlstd{= obj_tmb}\hlopt{$}\hlstd{par,} \hlkwc{objective} \hlstd{= obj_tmb}\hlopt{$}\hlstd{fn,}
                  \hlkwc{gradient} \hlstd{= obj_tmb}\hlopt{$}\hlstd{gr,} \hlkwc{hessian} \hlstd{= obj_tmb}\hlopt{$}\hlstd{he)}
\hlstd{mod_tmb}\hlopt{$}\hlstd{objective}
\end{alltt}
\begin{verbatim}
## [1] 168.5361
\end{verbatim}
\end{kframe}
\end{knitrout}
Note that passing the gradient and Hessian provided by {\tt{TMB}} to \texttt{nlminb} leads to the same minimum, i.e.~value of the nll function, here.

\end{enumerate}

\subsection{Basic nested model specification}
\label{sec:nested}

In the context of HMMs (and other statistical models), nested  models or models subject to certain parameter restrictions are commonly used. For example, it may be necessary to fix some parameters because of biological or physical constraints.  {\tt{TMB}} can be instructed to treat selected parameters as constants, or impose equality constraints on a set of parameters.
For the practical implementation, it is noteworthy that such parameter restrictions should be imposed on the working parameters.
However, it is also easily possible to impose restrictions on a natural parameter (e.g.~$\lambda$), and then identify the corresponding restriction on the working parameter (i.e.~$\log(\lambda)$). We illustrate a simple nested model specification by fixing $\lambda_1$ to one in our two-state Poisson HMM, the other parameter components correspond to the previous initial values. 
\begin{knitrout}
\definecolor{shadecolor}{rgb}{0.969, 0.969, 0.969}\color{fgcolor}\begin{kframe}
\begin{alltt}
\hlcom{# Get the previous values, and fix some}
\hlstd{fixed_par_lambda} \hlkwb{<-} \hlstd{lambda}
\hlstd{fixed_par_lambda[}\hlnum{1}\hlstd{]} \hlkwb{<-} \hlnum{1}
\end{alltt}
\end{kframe}
\end{knitrout}
We then transform these natural parameters into a set of working parameters.
\begin{knitrout}
\definecolor{shadecolor}{rgb}{0.969, 0.969, 0.969}\color{fgcolor}\begin{kframe}
\begin{alltt}
\hlcom{# Transform them into working parameters}
\hlstd{new_parameters} \hlkwb{<-} \hlkwd{pois.HMM.pn2pw}\hlstd{(}\hlkwc{m} \hlstd{= m,}
                                 \hlkwc{lambda} \hlstd{= fixed_par_lambda,}
                                 \hlkwc{gamma} \hlstd{= gamma)}
\end{alltt}
\end{kframe}
\end{knitrout}
For instructing {\tt{TMB}} to treat selected parameters as constants, the \texttt{map} argument of the \texttt{MakeADFun} has to be specified in addition to the usual arguments. The \texttt{map} argument is a list consisting factor-valued vectors which possess the same length as the working parameters and carry their names as well. 
The factor levels have to be unique for the regular parameters not subject to specific restrictions. If a parameter is fixed the corresponding entry of the \texttt{map} argument is filled with \texttt{NA}. In our example, this leads to:
\begin{knitrout}
\definecolor{shadecolor}{rgb}{0.969, 0.969, 0.969}\color{fgcolor}\begin{kframe}
\begin{alltt}
\hlstd{map} \hlkwb{<-} \hlkwd{list}\hlstd{(}\hlkwc{tlambda} \hlstd{=} \hlkwd{as.factor}\hlstd{(}\hlkwd{c}\hlstd{(}\hlnum{NA}\hlstd{,} \hlnum{1}\hlstd{)),}
            \hlkwc{tgamma} \hlstd{=} \hlkwd{as.factor}\hlstd{(}\hlkwd{c}\hlstd{(}\hlnum{2}\hlstd{,} \hlnum{3}\hlstd{)))}
\hlstd{fixed_par_obj_tmb} \hlkwb{<-} \hlkwd{MakeADFun}\hlstd{(TMB_data, new_parameters,}
                               \hlkwc{DLL} \hlstd{=} \hlstr{"poi_hmm"}\hlstd{,}
                               \hlkwc{silent} \hlstd{=} \hlnum{TRUE}\hlstd{,}
                               \hlkwc{map} \hlstd{= map)}
\end{alltt}
\end{kframe}
\end{knitrout}

It is noteworthy that more complex constraints are possible as well. For example, to impose equality constraints (such as $\gamma_{11} = \gamma_{22}$), the corresponding factor level has to be identical for the concerned entries. We refer to our GitHub page for details.

Last, estimation of the remaining model parameters and extraction of the results is achieved as before.
\begin{knitrout}
\definecolor{shadecolor}{rgb}{0.969, 0.969, 0.969}\color{fgcolor}\begin{kframe}
\begin{alltt}
\hlstd{fixed_par_mod_tmb} \hlkwb{<-} \hlkwd{nlminb}\hlstd{(}\hlkwc{start} \hlstd{= fixed_par_obj_tmb}\hlopt{$}\hlstd{par,}
                            \hlkwc{objective} \hlstd{= fixed_par_obj_tmb}\hlopt{$}\hlstd{fn,}
                            \hlkwc{gradient} \hlstd{= fixed_par_obj_tmb}\hlopt{$}\hlstd{gr,}
                            \hlkwc{hessian} \hlstd{= fixed_par_obj_tmb}\hlopt{$}\hlstd{he)}
\hlkwd{summary}\hlstd{(}\hlkwd{sdreport}\hlstd{(fixed_par_obj_tmb),} \hlstr{"report"}\hlstd{)}
\end{alltt}
\begin{verbatim}
##          Estimate Std. Error
## lambda 1.00000000 0.00000000
## lambda 5.50164872 0.30963641
## gamma  0.94561055 0.04791050
## gamma  0.02655944 0.02133283
## gamma  0.05438945 0.04791050
## gamma  0.97344056 0.02133283
## delta  0.32810136 0.22314460
## delta  0.67189864 0.22314460
\end{verbatim}
\end{kframe}
\end{knitrout}
Note that the standard error of $\lambda_1$ equals zero, because it is no longer considered a parameter and does not enter the optimization procedure.

\subsection{State inference and forecasting}
\label{sec:state_inference}

After estimating a HMM by the procedures illustrated in \autoref{sec:tmb_r}, it is possible to carry out a couple analyses that provide insight into the interpretation of the estimated model. These include, e.g., the so-called smoothing probabilities, which correspond to the probability of being in state $i$ at time $t$ for $i = 1,\ldots,m$, $t=1,\ldots,n$, given all observations. These probabilities can be obtained by
\begin{equation*}
\text{P}(C_t = i \vert X^{(n)} = x^{(n)}) = \frac{\alpha_t(i) \beta_t(i)}{L(\hat \btheta)},
\end{equation*}
where $\hat \btheta$ denotes the set of ML estimates. The derived smoothing probabilities then serve for determining the most probable state $i_t^*$ at time $t$ given the observations by
\begin{equation*}
i_t^* = \argmax_{i_t \in \{1, \ldots, m \}} \text{P}(C_t = i_t \vert X^{(n)} = x^{(n)}).
\end{equation*}
Furthermore, the Viterbi algorithm determines the overall most probable sequence of states $i_1^*, \ldots, i_T^*$, given the observations. This is achieved by evaluating
\begin{equation*}
(i_1^*, \ldots, i_n^*) = \argmax_{i_1, \ldots, i_n \in \{1, \ldots, m \}} \text{P}(C_1 = i_1, \ldots, C_n = i_n \vert X^{(n)} = x^{(n)}).
\end{equation*}
Other quantities of interest include the forecast distribution or $h$-step-ahead probabilities, which are obtained through
\begin{equation*}
\text{P}(X_{n+h} = x \vert X^{(n)} = x^{(n)}) = \frac{\balpha_n \bgamma^h \bcp(x) \bone'}{\balpha_n \bone'} = \bfphi_n \bgamma^h \bcp(x) \bone',
\end{equation*}
where $\bfphi_n = {\balpha_n} / {\balpha_n \bone'}$.

All the quantities shown above and the related algorithms for deriving them are described in detail in \citet[][Chapter 5]{zucchini}. In order to apply these algorithms, it is only necessary to extract the quantities required as input from a suitable \texttt{MakeADFun} object. Note that most algorithms rely on scaled versions of the forward- and backward-algorithm. This is illustrated in detail on GitHub. \autoref{fig:data_plot_tyt} shows the TYT data together with the conditional mean values linked to the most probable state inferred by the smoothing probabilities.

\begin{knitrout}
\definecolor{shadecolor}{rgb}{0.969, 0.969, 0.969}\color{fgcolor}\begin{figure}[htb]

{\centering \includegraphics[width=\maxwidth]{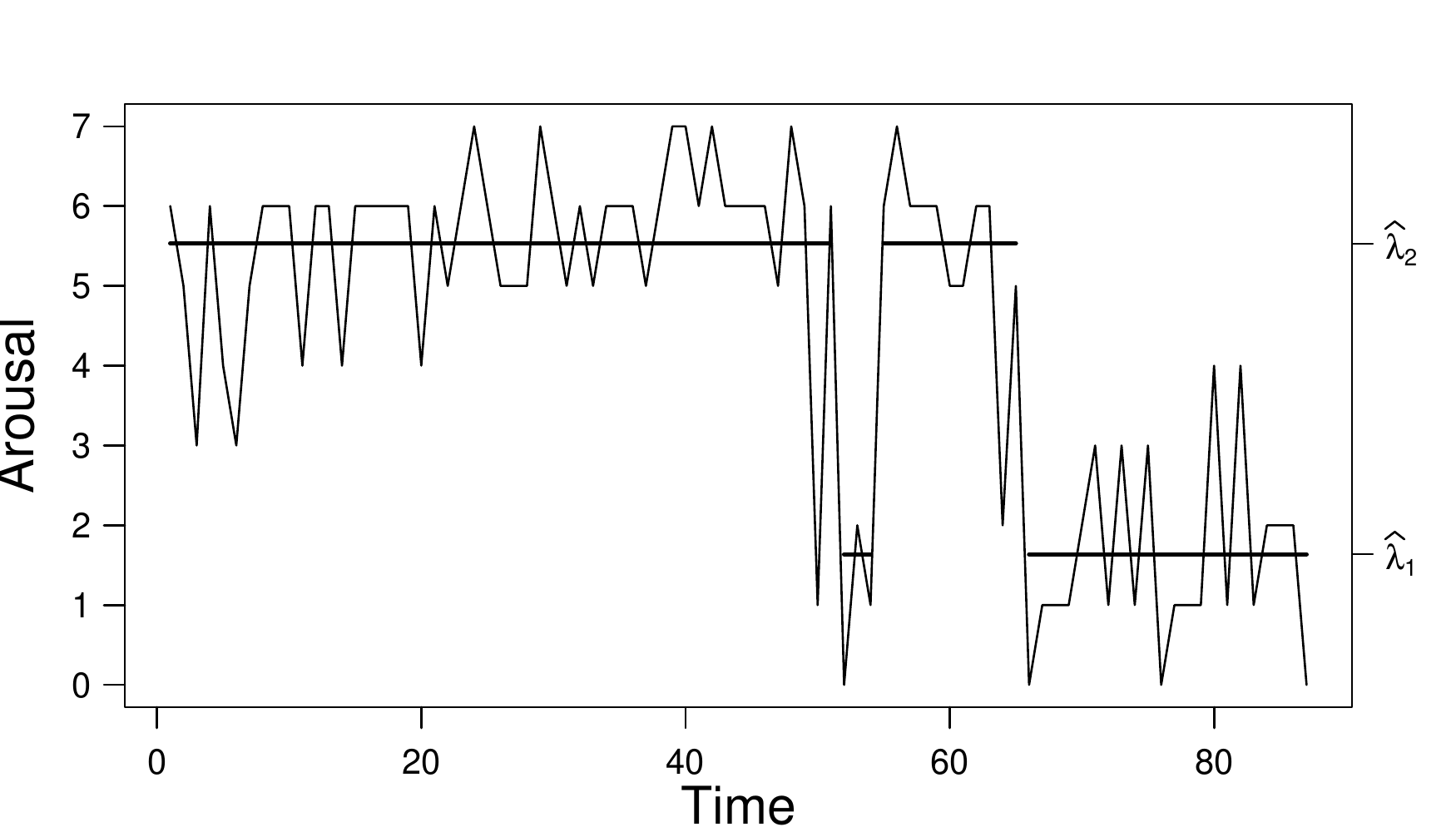} 

}

\caption{Plot of the TYT data. The solid horizontal lines correspond to the conditional mean of the inferred state at each time. See \autoref{table:tinn_cis} for the values of $\widehat{\lambda}_i$.}\label{fig:data_plot_tyt}
\end{figure}

\end{knitrout}

\section{Confidence intervals}
\label{sec:confint}

A common approach for deriving confidence intervals (CIs) for the estimated parameters of statistical models bases on finite-difference approximations of the Hessian. This technique is, however, not suited for most HMMs due to computational difficulties, as already pointed out by \citet{visser}. The same authors suggest likelihood profile CIs or bootstrap-based CIs as potentially better alternatives. Despite the potentially high computational load, bootstrap-based CIs have become an established method in the context of HMMs \citep{bulla, zucchini} and found widespread application by practitioners.

In this section we illustrate how CIs based on the Hessian, likelihood profiling, and the bootstrap can be efficiently implemented by integrating {\tt{TMB}}.
This permits in particular to obtain Hessian based and likelihood profile based CIs at very low computational cost. For simplicity, we illustrate our procedures by means of the parameter $\lambda_2$ of our two-state Poisson HMM. We will further address the resulting CIs for $\bgamma$ and $\bflambda$ and performance-related aspects in \autoref{sec:application_datasets}.

\subsection{Wald-type confidence intervals based on the Hessian}
\label{sec:hessian}

Since the negative log-likelihood function of HMMs typically depends on the working parameters, evaluation of the Hessian in the optimum found by numerical optimization only serves for inference about the working parameters. From a practical perspective, however, inference about the natural parameters usually is of interest. As the Hessian $\nabla^2\log L(\{\hat{\bct}, \hat{\bfeta}\})$ refers to the working parameters $\{\bct, \bfeta\}$, the delta method is suitable to obtain an estimate of the covariance matrix of $\{\hat{\bgamma}, \hat{\bflambda}\}$ by
\begin{equation}
\Sigma_{\hat{\bgamma}, \hat{\bflambda}} = - \nabla g(\hat{\bct}, \hat{\bfeta})\left(\nabla^2\log L(\hat{\bct}, \hat{\bfeta})\right)^{-1}\nabla g(\hat{\bct}, \hat{\bfeta})^\prime,
\label{eq:deltamethod}
\end{equation}
with $\{\hat{\bgamma}, \hat{\bflambda}\} = g(\hat{\bct}, \hat{\bfeta})$ as defined in \autoref{sec:hmm_repar}.
From a user's perspective, it is highly convenient that the entire right-hand side of \autoref{eq:deltamethod} can be directly computed via automatic differentiation in \texttt{TMB}. Moreover, it is particularly noteworthy that the standard errors of derived parameters can be calculated by the delta-method similarly. For example, the stationary distribution $\bfdelta$ is a function of $\bgamma$ in our case, and \texttt{TMB} provides a straightforward way to obtain standard errors of $\bfdelta$. This is achieved by first defining $\bfdelta$ inside the {\tt{C++}} file \texttt{poi\_hmm.cpp} (or, in our implementation, the related \texttt{utils.cpp}, which gathers auxiliary functions). Secondly, it is necessary to call \texttt{ADREPORT} on $\bfdelta$ within the \texttt{poi\_hmm.cpp} file. To display the resulting estimates and corresponding standard errors in {\tt{R}}, one can rely on the command shown previously in \autoref{sec:tmb_r}.

Subsequently, Wald-type confidence intervals \citep{wald} follow in the usual manner. For example, the $(1 - \alpha) \%$ CI for $\lambda_1$ is given by $\lambda_1 \pm z_{1-\alpha/2} * \sigma_{\lambda_1}$ where $z_{x}$ is the $x$-percentile of the standard normal distribution, and $\sigma_{\lambda_1}$ is the standard error of $\lambda_1$ obtained via the delta method. This part is easily implemented in {\tt{R}}. We illustrate the calculation of these CIs for our two-state Poisson HMM on GitHub.

Finally, note that the reliability of Wald-type CIs may suffer from a singular Fisher information matrix, which can occur for many different types of statistical models, including HMMs. This also jeopardizes the validity of AIC and BIC criteria. For further details on this topic, see, e.g., \citet{drton}.

\subsection{Likelihood profile based confidence intervals}
\label{sec:likelihood}

The Hessian based CIs presented above rely on asymptotic normality of the ML estimator. Properties of the ML estimator may, however, change in small samples. Moreover, symmetric CIs may not be suitable if the ML estimator lies close to a boundary of the parameter space. This occurs, e.g., when states are highly persistent, which leads to entries close to one in the TPM. An alternative approach to construct CIs bases on the so-called profile likelihood \citep[see, e.g.,][]{venzon, meeker}, which has also shown a satisfactory performance in the context of HMMs \citep{visser}.

In the following, we illustrate the principle of likelihood profile based CIs by the example of the parameter $\lambda_2$ in our two-state Poisson HMM. The underlying basic idea is to identify those values of our parameter of interest $\lambda_2$ in the neighborhood of $\hat \lambda_2$ that lead to a significant change in the log-likelihood, whereby the other parameters (i.e.~$\bgamma$, $\lambda_1$) are considered nuisance parameters \citep{meeker}. The term ``nuisance parameters" means that these parameters need to be re-estimated (by maximizing the likelihood) for any fixed value of $\lambda_2$ different to $\hat \lambda_2$. That is, the profile likelihood of $\lambda_2$ is defined as
\begin{equation*}
L_p(\lambda_2) = \max_{\bgamma, \lambda_1} L(\bgamma, \bflambda).
\end{equation*}

In order to construct profile likelihood-based CIs, let $\{\hat{\bgamma}, \hat{\bflambda}\}$ denote the ML estimate for our HMM computed as described in \autoref{sec:tmb_r}. Evaluation of the log-likelihood function in this point results in the value $\log L(\{\hat{\bgamma}, \hat{\bfdelta}\})$. The deviation of the likelihood of the ML estimate and the profile likelihood in the point $\lambda_2^p$ is then captured by the following likelihood ratio:
\begin{equation}
R_p(\lambda_2) = -2 \left[ \log(L_p(\lambda_2)) - \log(L(\hat{\bgamma}, \hat{\bflambda}))\right].
\label{eq:profileLR}
\end{equation}
As described above, the log-likelihood $\log(L_p(\lambda_2))$ results from re-estimating the two-state HMM with fixed parameter $\lambda_2$. Therefore, this model effectively corresponds to a nested model of the full model with ML estimate $\hat{\bgamma}, \hat{\bflambda}$. Consequently, $R_p$ asymptotically follows a $\chi^{2}$ distribution with one degree of freedom - the difference in degrees of freedom between the two models. Based on this, a CI for $\lambda_2$ can be derived by evaluating $R_p$ at many different values of $\lambda_2^p$ and determining when the resulting value of $R_p$ becomes ``too extreme". That is, for a given $\alpha$, one needs to calculate the $1-\alpha$ quantile of the $\chi^{2}_{1}$ distribution (e.g., 3.841 for $\alpha = 5\%$). The CI at level $1-\alpha$ for the parameter $\lambda_2$ is then given by 
\begin{equation}
\left\{\lambda_2: R_p(\lambda_2)  < \chi^{2}_{1, (1-\alpha)}\right\}.
\label{eq:profileCI}
\end{equation}

\bigskip

For simplicity, the principles of likelihood profiling shown above rely on the natural parameters.
Our nll function is, however, parametrized in terms of and optimized with respect to the working parameters.
In practice, this aspect is easy to deal with.
Once a profile CI for the working parameter (here $\eta_2$) has been obtained following the procedure above, the corresponding CI for the natural parameter $\lambda_2$ results directly from transforming the upper and lower boundary of the CI for $\eta_2$ by the one-to-one transformation $\lambda_2 = \exp(\eta_2)$.
For further details on the invariance of likelihood-based CIs to parameter transformations, {\hl see, e.g.,} \citet{meeker}.

{\tt{TMB}} provides an easy way to profiling through the function {\tt{tmbprofile}}, which requires several inputs.
First, the \texttt{MakeADFun} object called \texttt{obj\_tmb} from our two-state Poisson HMM.
Secondly, the position of the (working) parameter to be profiled via the \texttt{name} argument.
This position refers to the position in the parameter vector \texttt{obj\_tmb\$par}.
Moreover, here the optional \texttt{trace} argument indicates how much information on the optimization is displayed.
The following commands permit to profile the second working parameter $\eta_2 = \log(\lambda_2)$.

\begin{knitrout}
\definecolor{shadecolor}{rgb}{0.969, 0.969, 0.969}\color{fgcolor}\begin{kframe}
\begin{alltt}
\hlcom{# Parameters and covariates}
\hlstd{m} \hlkwb{<-} \hlstd{M_LIST_TINN}
\hlkwa{if} \hlstd{(m} \hlopt{==} \hlnum{1}\hlstd{) \{}
  \hlstd{gamma} \hlkwb{<-} \hlkwd{matrix}\hlstd{(}\hlnum{1}\hlstd{)}
\hlstd{\}} \hlkwa{else} \hlstd{\{}
  \hlstd{gamma} \hlkwb{<-} \hlkwd{matrix}\hlstd{(}\hlnum{0.2} \hlopt{/} \hlstd{(m} \hlopt{-} \hlnum{1}\hlstd{),} \hlkwc{nrow} \hlstd{= m,} \hlkwc{ncol} \hlstd{= m)}
  \hlkwd{diag}\hlstd{(gamma)} \hlkwb{<-} \hlnum{0.8}
\hlstd{\}}
\hlstd{lambda} \hlkwb{<-} \hlkwd{seq}\hlstd{(}\hlkwc{from} \hlstd{=} \hlkwd{quantile}\hlstd{(tinn_data,} \hlnum{0.1}\hlstd{),}
              \hlkwc{to} \hlstd{=} \hlkwd{quantile}\hlstd{(tinn_data,} \hlnum{0.9}\hlstd{),}
              \hlkwc{length.out} \hlstd{= m)}
\hlstd{delta} \hlkwb{<-} \hlkwd{stat.dist}\hlstd{(gamma)}

\hlcom{# Parameters & covariates for TMB}
\hlstd{working_params} \hlkwb{<-} \hlkwd{pois.HMM.pn2pw}\hlstd{(m, lambda, gamma)}
\hlstd{TMB_data} \hlkwb{<-} \hlkwd{list}\hlstd{(}\hlkwc{x} \hlstd{= tinn_data,} \hlkwc{m} \hlstd{= m)}

\hlcom{# Estimation}
\hlstd{tmb_gh} \hlkwb{<-} \hlkwd{TMB.estimate}\hlstd{(}\hlkwc{TMB_data} \hlstd{= TMB_data,}
                       \hlkwc{parameters} \hlstd{= working_params,}
                       \hlkwc{gradient} \hlstd{=} \hlnum{TRUE}\hlstd{,}
                       \hlkwc{hessian} \hlstd{=} \hlnum{TRUE}\hlstd{)}

\hlcom{# Profile}
\hlstd{profile} \hlkwb{<-} \hlkwd{tmbprofile}\hlstd{(}\hlkwc{obj} \hlstd{= tmb_gh}\hlopt{$}\hlstd{obj,}
                      \hlkwc{name} \hlstd{=} \hlnum{2}\hlstd{,}
                      \hlkwc{trace} \hlstd{=} \hlnum{FALSE}\hlstd{)}
\hlkwd{par}\hlstd{(}\hlkwc{mgp} \hlstd{=} \hlkwd{c}\hlstd{(}\hlnum{2}\hlstd{,} \hlnum{0.5}\hlstd{,} \hlnum{0}\hlstd{),} \hlkwc{mar} \hlstd{=} \hlkwd{c}\hlstd{(}\hlnum{3}\hlstd{,} \hlnum{3}\hlstd{,} \hlnum{2.5}\hlstd{,} \hlnum{1}\hlstd{),}
    \hlkwc{cex.lab} \hlstd{=} \hlnum{1.5}\hlstd{)}
\hlkwd{plot}\hlstd{(profile,} \hlkwc{level} \hlstd{=} \hlnum{0.95}\hlstd{,}
     \hlkwc{xlab} \hlstd{=} \hlkwd{expression}\hlstd{(eta[}\hlnum{2}\hlstd{]),}
     \hlkwc{ylab} \hlstd{=} \hlstr{"nll"}\hlstd{)}
\end{alltt}
\end{kframe}\begin{figure}[htb]

{\centering \includegraphics[width=\maxwidth]{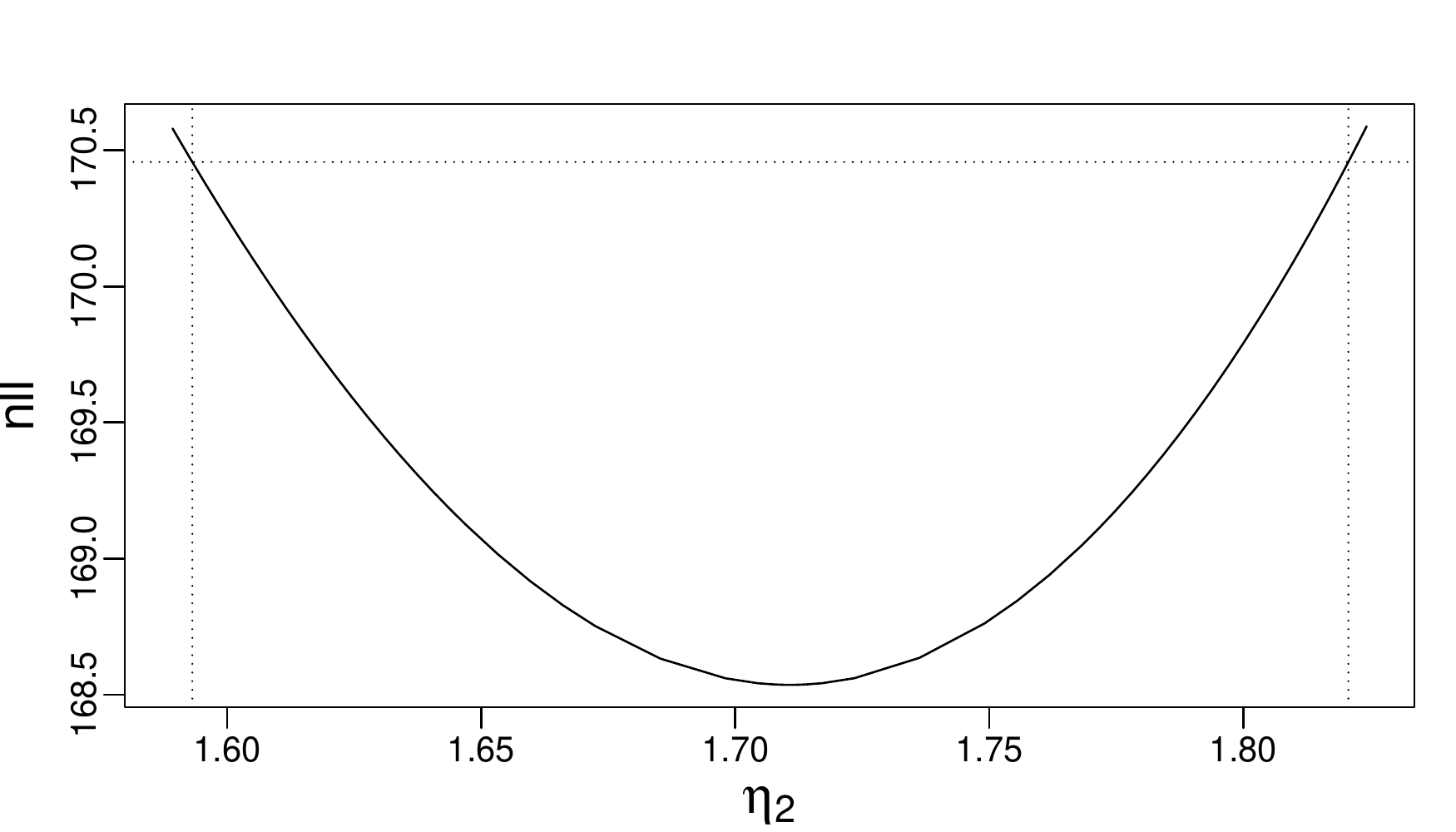} 

}

\caption[Profile likelihood plot]{Profile likelihood plot. This figure shows the profile nll as function of the working parameter $\eta_2$. The vertical and horizontal lines correspond to the boundaries of the confidence interval and the critical value of the nll, respectively.}\label{fig:profile_plot}
\end{figure}

\end{knitrout}

Furthermore, \autoref{fig:profile_plot} obtained via the \texttt{plot} function shows the resulting profile nll as function of the working parameter $\eta_2$. The vertical and horizontal lines correspond to the boundaries of the 95\% CI and the critical value of the nll derived from \autoref{eq:profileCI}, respectively. The CI for $\eta_2$ can directly be extracted via the function \texttt{confint}:
\begin{knitrout}
\definecolor{shadecolor}{rgb}{0.969, 0.969, 0.969}\color{fgcolor}\begin{kframe}
\begin{alltt}
\hlcom{# Confidence interval of tlambda}
\hlkwd{confint}\hlstd{(profile,} \hlkwc{level} \hlstd{=} \hlnum{0.95}\hlstd{)}
\end{alltt}
\begin{verbatim}
##            lower    upper
## tlambda 1.593141 1.820641
\end{verbatim}
\end{kframe}
\end{knitrout}
The corresponding CI for $\lambda_2$ the follows from:
\begin{knitrout}
\definecolor{shadecolor}{rgb}{0.969, 0.969, 0.969}\color{fgcolor}\begin{kframe}
\begin{alltt}
\hlcom{# Confidence interval of lambda}
\hlkwd{exp}\hlstd{(}\hlkwd{confint}\hlstd{(profile,} \hlkwc{level} \hlstd{=} \hlnum{0.95}\hlstd{))}
\end{alltt}
\begin{verbatim}
##            lower    upper
## tlambda 4.919178 6.175815
\end{verbatim}
\end{kframe}
\end{knitrout}
While simple linear combinations of variables can be profiled through the argument \texttt{lincomb} in the \texttt{tmbprofile} function, this is not possible for more complex functions of the parameters. This includes the stationary distribution $\bfdelta$, for which CIs cannot be obtained by this method.

Note that the function \texttt{tmbprofile} carries out several optimization procedures internally for calculating profile CIs. If this approach fails, or one prefers a specific optimization routine, the necessary steps for profiling can also be implemented by the user. To do so, it would be - roughly speaking - necessary to compute $R_p(\eta_2)$ for a sufficient number of $\eta_2$ values to achieve the desired precision.

{\hl
Last, it is noteworthy that calculating profile CIs for the elements of the TPM is not trivial.
Since all elements in each row of the TPM depend on each other, the definition of the profile log-likelihood is already problematic.
We apply a rather simplistic approach and compute CIs for each of the working parameters $\tau_{i,j}, i \neq j$.
The resulting upper and lower bounds of the profile CI are then transformed to their natural counterparts.
However, other approaches may lead to improved results, see, e.g., \citet{fischer}.
Note that this approach seems to produce biased results for models with more than two states due to the dependence between row elements of the TPM.
Hence, we advise to treat profile CIs for the TPM with care, in particular for HMMs with more than two states.}


\subsection{Bootstrap-based confidence intervals}
\label{sec:bootstrapping}

The last approach for deriving CIs is the bootstrap, which is frequently applied by many practitioner. \citet{efron} describe the underlying concepts of the bootstrap in their seminal manuscript. Many different bootstrap techniques have evolved since then, leading to an extensive treatment of this subject in the scientific literature.

A thorough overview of this subject would go beyond the scope of this paper. As pointed out by \citet{hardle}, the so-called parametric bootstrap is suitable in the context of time series models. For further details on the bootstrap for HMMs including the implementation of a parametric bootstrap, we refer to \citet[][Ch.~3, pp.~56-60]{zucchini}.

Basically all versions of the bootstrap have in common that some kind of re-sampling procedure needs to be carried out first. Secondly, the model of interest is re-estimated for each of the re-sampled data sets. A natural way to accelerate the second part consists in the use of \texttt{TMB} for the model estimation by means of the procedures presented in \autoref{sec:tmb_r}. Our GitHub page contains a detailed example illustrating the implementation of a parametric percentile bootstrap for our two-state Poisson HMM.

\section{Application to different data sets}
\label{sec:application_datasets}

This section aims to demonstrate the performance of \texttt{TMB} by means of a couple of practical examples that differ in terms of the number of observations and model complexity. These examples include the TYT data shown above, a data set of fetal lamb movements, and simulated data sets. For the performance comparisons, the focus lies on computational speed and the reliability of confidence intervals. The {\tt{R}} scripts necessary for this section may serve interested users for investigating their own HMM setting, and are all available on GitHub.

\subsection{TYT data}
\label{sec:tyt_data}

We begin by investigating the speed of five approaches for parameter estimation: one without the usage of {\tt{TMB}}, and four with {\tt{TMB}}. In the following, $DM$ denotes direct maximization of the log-likelihood through the optimization function \texttt{nlminb} without {\tt{TMB}}. Furthermore, $TMB_0$, $TMB_H$, $TMB_G$, and $TMB_{GH}$ denote direct maximization with {\tt{TMB}} without using the gradient and Hessian provided by {\tt{TMB}}, with the Hessian, with the gradient, and with both gradient and Hessian, respectively.

As a preliminary reliability check of our IT infrastructure and setup, we timed the fitting of our two-state HMM to the TYT data with the help of the {\tt{microbenchmark}} package \citep{R-microbenchmark}. For this data set, all five approaches converged to the same optimum and parameter estimates, apart from minor variations typical for numerical optimization (see \autoref{table:2-state-tinn-estimates}).

\begin{table}[ht]
\centering
\begin{tabular}{cccccc}
  \hline
Par. & \textit{${DM}$} & \textit{${TMB_0}$} & \textit{${TMB_G}$} & \textit{${TMB_H}$} & \textit{${TMB_{GH}}$} \\ 
  \hline
$\lambda_{1}$ & 1.636410931 & 1.636410932 & 1.636410933 & 1.636410932 & 1.636410997 \\ 
  $\lambda_{2}$ & 5.533095962 & 5.533095962 & 5.533095957 & 5.533095962 & 5.533095759 \\ 
  $\gamma_{11}$ & 0.949802041 & 0.949802041 & 0.949802042 & 0.949802041 & 0.949802094 \\ 
  $\gamma_{12}$ & 0.050197959 & 0.050197959 & 0.050197958 & 0.050197959 & 0.050197906 \\ 
  $\gamma_{21}$ & 0.025922044 & 0.025922044 & 0.025922044 & 0.025922044 & 0.025922038 \\ 
  $\gamma_{22}$ & 0.974077956 & 0.974077956 & 0.974077956 & 0.974077956 & 0.974077962 \\ 
  $\delta_{1}$ & 0.340541816 & 0.340541816 & 0.340541819 & 0.340541816 & 0.340541999 \\ 
  $\delta_{2}$ & 0.659458184 & 0.659458184 & 0.659458181 & 0.659458184 & 0.659458001 \\ 
  nll & 168.536055869 & 168.536055869 & 168.536055869 & 168.536055869 & 168.536055869 \\ 
   \hline
\end{tabular}
\caption{Parameter estimates and corresponding nll of the two-state Poisson HMM with and without using {\tt TMB} obtained for the TYT data.} 
\label{table:2-state-tinn-estimates}
\end{table}

\autoref{table:speed-consistency-tinn} shows the resulting average time required for the parameter estimation and the number of iterations needed by each approach, measured over $200$ replications. The results show that the use of {\tt{TMB}} significantly accelerates parameter estimation in comparison with $DM$. The most substantial acceleration is achieved by $TMB_G$, underlining the benefit of using the gradient provided by {\tt{TMB}}. Moreover, $TMB_{GH}$ requires fewer iterations than the other approaches. However, the evaluation of the Hessian seems to increase the computational burden.     

\begin{table}[ht]
\centering
\begin{tabular}{lccccc}
  \hline
 & \textit{${DM}$} & \textit{${TMB_0}$} & \textit{${TMB_G}$} & \textit{${TMB_H}$} & \textit{${TMB_{GH}}$} \\ 
  \hline
Time (ms) & 36.5 & 1.66 & 0.824 & 1.64 & 2.62 \\ 
   &  (36.2, 36.9)  &  (1.6, 1.72)  &  (0.822, 0.826)  &  (1.6, 1.68)  &  (2.59, 2.65)  \\ 
  Iterations & 13 & 13 & 13 & 13 & 7 \\ 
   \hline
\end{tabular}
\caption{Average duration (in milliseconds) together with 95\% CI and number of iterations required for fitting a two-state Poisson HMM to the TYT data. The CIs are of Wald-type and base on the standard error of the mean derived from 200 replications.} 
\label{table:speed-consistency-tinn}
\end{table}

Next, we verified the reproducibility of the acceleration by {\tt{TMB}} in a parametric bootstrap setting. More specifically, we simulated $200$ bootstrap samples from the model estimated on the TYT data. Then, we re-estimated the same model by our five approaches and derived acceleration ratios (with $DM$ as reference approach) and their corresponding percentile CIs. As shown in \autoref{table:speed-tinn}, all acceleration ratios take values significantly larger than one, whether the gradient and Hessian are passed from {\tt{TMB}} or not. In addition, the findings from the single TYT data set are confirmed, with $TMB_G$ providing the most substantial acceleration and $TMB_{GH}$ reducing the number of iterations. This underlines that two factors are sources of the acceleration in \autoref{table:speed-tinn}: the use of {\tt{C++}} code on the one hand, and computation of the gradient and/or Hessian by {\tt TMB} on the other hand.  

\begin{table}[ht]
\centering
\begin{tabular}{lcccc}
  \hline
 & \textit{${TMB_0}$} & \textit{${TMB_G}$} & \textit{${TMB_H}$} & \textit{${TMB_{GH}}$} \\ 
  \hline
Acceleration ratio & 21.8 & 41.8 & 21.8 & 16.7 \\ 
   &  (20.7, 23)  &  (37.4, 46.8)  &  (20.8, 23.3)  &  (12.9, 22.1)  \\ 
  Iterations & 12.6 & 12.6 & 12.6 & 5.07 \\ 
   &  (9, 18)  &  (9, 19)  &  (9, 18)  &  (4, 8.03)  \\ 
   \hline
\end{tabular}
\caption{Acceleration and iterations for the TYT data. The top lines show the acceleration ratios together with 95\% percentile bootstrap CIs when using {\tt TMB} in a bootstrap setting with 200 bootstrap samples. The bottom lines display the corresponding values for the number of iterations. {\hl The average parameter estimation duration without {\tt TMB} is 37 milliseconds.}} 
\label{table:speed-tinn}
\end{table}

In order to obtain reliable results, we excluded all those bootstrap samples with a simulated state sequence not sojourning in each state of the underlying model at least once. This is necessary because such a constellation almost certainly leads to convergence problems on the one hand. On the other hand, even if the estimation algorithms converge, the estimated models are usually degenerate because of the lack of identifiability. Furthermore, for some very rare bootstrap samples, one or several of the estimation algorithms did not converge properly. In such cases, we discarded the results, generated an additional bootstrap sample, and re-ran the parameter estimation. Convergence problems mainly occurred due to $TMB_0$ and $TMB_H$ failing. Therefore, we recommend passing at least the gradient when optimizing with {\tt{TMB}} for increased stability.

Last, we investigate CIs obtained for the TYT data by the three different methods described in \autoref{sec:confint}, $TMB_{GH}$ served as the sole estimation approach. The columns to the left in \autoref{table:tinn_cis} show the parameter estimates and the three types of 95\% CIs obtained using the Hessian, likelihood profiling, and bootstrapping. For this data set, no major differences between the different CIs are visible. Furthermore, we assessed the accuracy of the different CIs by computing coverage probabilities, which are shown in the last three columns of \autoref{table:tinn_cis}. For calculating these coverage probabilities, we used a Monte Carlo setting similar to the one described above. Samples that possessed state sequences not visiting all states or samples for which the estimation algorithm did not converge were replaced. Moreover, we also simulated a replacement sequence when the profile likelihood method failed to converge on any bound to ensure comparability of the results. The results, shown on the right of \autoref{table:tinn_cis} indicate that all methods provide comparably reliable CI estimates, and neither outperforms the other for all parameters. For the Wald-type CIs, the coverage probabilities almost reach 100\% for $\gamma_{22}$, $\gamma_{21}$, but lie comparably low for both the other elements of the TPM and $\delta_1$, $\delta_2$. However, profile likelihood-based CIs also take values above 95\% for all elements of the TPM, and the coverage probabilities for bootstrap CIs are all above 95\%.
{\hl Altogether, it appears that the reliability of CIs should be investigated carefully when fitting HMMs to very short sequences of observations.}

\begin{table}[ht]
\centering
\begin{tabular}{ccccccccccc}
  && \multicolumn{2}{c}{Wald-type CI}& \multicolumn{2}{c}{Profile CI}& \multicolumn{2}{c}{Bootstrap CI}& \multicolumn{3}{c}{Coverage prob. (\%)}\\ \hline
Par. & Est. & L. & U. & L. & U. & L. & U. & Wald & Profile & Bootst. \\ 
  \hline
$\lambda_{1}$ & 1.64 & 1.09 & 2.18 & 1.15 & 2.23 & 0.95 & 2.93 & 92.7 & 94.9 & 98.8 \\ 
  $\lambda_{2}$ & 5.53 & 4.91 & 6.16 & 4.92 & 6.18 & 4.88 & 6.31 & 93.5 & 93.5 & 98.4 \\ 
  $\gamma_{11}$ & 0.95 & 0.86 & 1.00 & 0.82 & 1.00 & 0.54 & 0.99 & 89.7 & 97.3 & 96.5 \\ 
  $\gamma_{12}$ & 0.05 & 0.00 & 0.14 & 0.00 & 0.18 & 0.01 & 0.46 & 89.7 & 97.3 & 96.5 \\ 
  $\gamma_{21}$ & 0.03 & 0.00 & 0.07 & 0.00 & 0.09 & 0.01 & 0.14 & 100.0 & 95.6 & 95.5 \\ 
  $\gamma_{22}$ & 0.97 & 0.93 & 1.00 & 0.91 & 1.00 & 0.86 & 0.99 & 100.0 & 95.6 & 95.5 \\ 
  $\delta_{1}$ & 0.34 & 0.00 & 0.79 &  &  & 0.07 & 0.79 & 86.2 &  & 95.8 \\ 
  $\delta_{2}$ & 0.66 & 0.21 & 1.00 &  &  & 0.21 & 0.93 & 86.2 &  & 95.8 \\ 
   \hline
\end{tabular}
\caption{CIs for the TYT dataset. From left to right, the columns contain: the parameter name, parameter estimate, and lower (L.) and upper (U.) bound of the corresponding 95\% CI derived via the Hessian provided by {\tt TMB}, likelihood profiling, and percentile bootstrap. Then follow coverage probabilities derived for these three methods in a Monte-Carlo study.} 
\label{table:tinn_cis}
\end{table}

\subsection{Lamb data}
\label{sec:lamb_data}

The well-known data set presented in \citet{leroux} serves as second example. This data set consists of the number of movements by a fetal lamb observed through ultrasound during 240 consecutive 5-second intervals, as shown in \autoref{fig:data_plot_lamb}. Since the results reported by \citet{leroux} show that a two-state model is preferred by the BIC, we focus on this model only here - although other choices would be possible, e.g.~the AIC selects a three-state model.

\begin{knitrout}
\definecolor{shadecolor}{rgb}{0.969, 0.969, 0.969}\color{fgcolor}\begin{figure}[htb]

{\centering \includegraphics[width=\maxwidth]{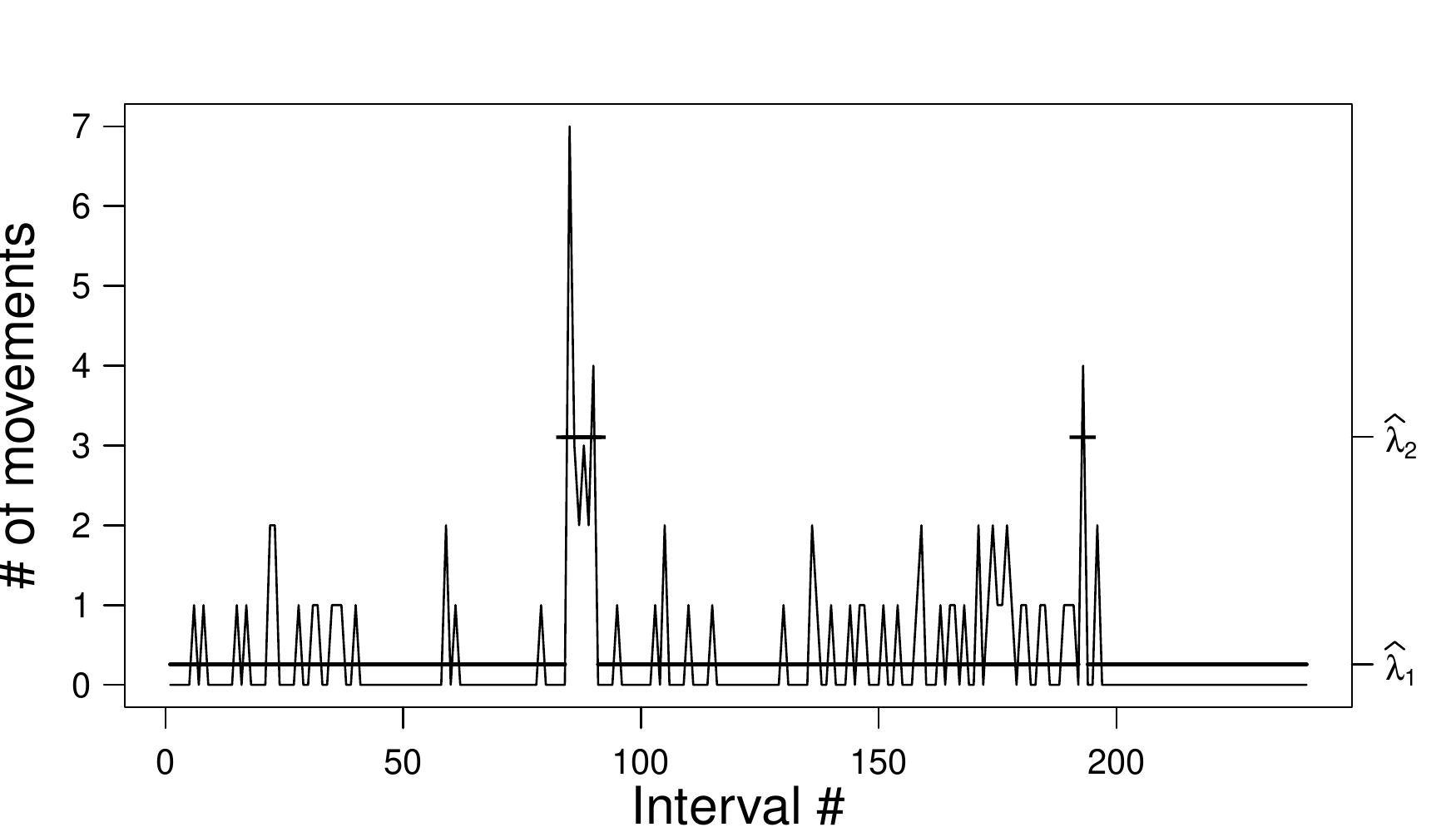} 

}

\caption{Plot of the lamb data. The solid horizontal lines correspond to the conditional mean of the inferred state at each time. See \autoref{table:lamb_cis} for the values of $\widehat{\lambda}_i$.}\label{fig:data_plot_lamb}
\end{figure}

\end{knitrout}

We selected this data set for several reasons. First, the number of observations is larger than for the TYT data (but still comparably low). Secondly, according to the results of \citet{leroux}, the first state largely dominates the data generating process, whereas the second state is not very persistent and linked to only a few observations. Thirdly, the conditional means of the two states are not very different. The latter two aspects qualify this data as a ``non-textbook example".

Similar to the TYT data, all estimation algorithms converged to the same minimum of the nll, and provided almost identical parameter estimates on the original data set. A bootstrap experiment similar to the one described above for the TYT data led to comparable results, as shown in \autoref{table:speed-lamb}. The highest acceleration is achieved by $TMB_G$, whereas $TMB_{GH}$ achieves the lowest acceleration despite requiring a lower number of iterations than the other approaches. However, all ratios lie above those obtained for the TYT data, indicating an increased benefit of using {\tt TMB} with an increasing number of observations.

\begin{table}[ht]
\centering
\begin{tabular}{lcccc}
  \hline
 & \textit{${TMB_0}$} & \textit{${TMB_G}$} & \textit{${TMB_H}$} & \textit{${TMB_{GH}}$} \\ 
  \hline
Acceleration ratio & 26.8 & 49.4 & 26.7 & 14.8 \\ 
   &  (25, 29.1)  &  (44.2, 64.8)  &  (25.3, 29)  &  (9.17, 20.5)  \\ 
  Iterations & 13.3 & 13.3 & 13.3 & 6.28 \\ 
   &  (8.98, 33)  &  (8.98, 32)  &  (8.98, 33)  &  (4, 22)  \\ 
   \hline
\end{tabular}
\caption{Acceleration and iterations for the lamb data. The top lines show the acceleration ratios together with 95\% percentile bootstrap CIs when using {\tt TMB} in a bootstrap setting with 200 bootstrap samples. The bottom lines display the corresponding values for the number of iterations. {\hl The average parameter estimation duration without {\tt TMB} is 76 milliseconds.}} 
\label{table:speed-lamb}
\end{table}

Next, \autoref{table:lamb_cis} shows parameter estimates and corresponding CIs in the columns to the left. Our estimates confirm the results of \cite{leroux}: the second state is not very persistent, and the conditional means $\lambda_1$ and $\lambda_2$ do not lie very far from each other. Concerning the CIs resulting from our three approaches, it is noticeable that the bootstrap CIs for the elements of the TPM are larger than those obtained by the other two approaches. The coverage probabilities presented in the columns to the right of \autoref{table:lamb_cis} show similar patterns as observed for the TYT data. Wald-type CIs show certain deviations from 95\% for the parameters related to the hidden state sequence, and the same is true for the profile CIs. Bootstrap CIs seem to be slightly too large for most parameters, leading to values greater than 95\%.
{\hl As for the TYT data, we recommend to be aware of the limited reliability of CIs for short sequences of observations.}

\begin{table}[ht]
\centering
\begin{tabular}{ccccccccccc}
  && \multicolumn{2}{c}{Wald-type CI}& \multicolumn{2}{c}{Profile CI}& \multicolumn{2}{c}{Bootstrap CI}& \multicolumn{3}{c}{Coverage prob. (\%)}\\ \hline
Par. & Est. & L. & U. & L. & U. & L. & U. & Wald & Profile & Bootst. \\ 
  \hline
$\lambda_{1}$ & 0.26 & 0.18 & 0.34 & 0.15 & 0.33 & 0.08 & 0.33 & 95.0 & 96.2 & 95.7 \\ 
  $\lambda_{2}$ & 3.11 & 1.11 & 5.12 & 1.27 & 4.95 & 0.50 & 5.32 & 93.3 & 94.6 & 97.7 \\ 
  $\gamma_{11}$ & 0.99 & 0.97 & 1.00 & 0.93 & 1.00 & 0.71 & 1.00 & 99.9 & 96.1 & 97.9 \\ 
  $\gamma_{12}$ & 0.01 & 0.00 & 0.03 & 0.00 & 0.07 & 0.00 & 0.29 & 99.9 & 96.1 & 97.9 \\ 
  $\gamma_{21}$ & 0.31 & 0.00 & 0.67 & 0.04 & 0.68 & 0.10 & 1.00 & 93.3 & 97.6 & 98.8 \\ 
  $\gamma_{22}$ & 0.69 & 0.33 & 1.00 & 0.32 & 0.96 & 0.00 & 0.90 & 93.3 & 97.6 & 98.8 \\ 
  $\delta_{1}$ & 0.96 & 0.90 & 1.00 &  &  & 0.61 & 0.99 & 98.2 &  & 98.9 \\ 
  $\delta_{2}$ & 0.04 & 0.00 & 0.10 &  &  & 0.01 & 0.39 & 98.2 &  & 98.9 \\ 
   \hline
\end{tabular}
\caption{CIs for the lamb dataset. From left to right, the columns contain: the parameter name, parameter estimate, and lower (L.) and upper (U.) bound of the corresponding 95\% CI derived via the Hessian provided by {\tt TMB}, likelihood profiling, and percentile bootstrap. Then follow coverage probabilities derived for these three methods in a Monte-Carlo study.} 
\label{table:lamb_cis}
\end{table}

On a minor note, some non-negligible differences can be noted when comparing our estimation results to those reported by \citet{leroux}. The reasons for this are difficult to determine, but some likely explanations are given in the following. First, differences in the parameter estimates may result e.g.~from the optimizing algorithms used and related setting (e.g.~convergence criterion, number of steps, optimization routines used in 1992,\dots). Moreover, \citet{leroux} seem to base their calculations on an altered likelihood, which is reduced by removing the constant term $\sum_{i=1}^{T} \log(x_{i}!)$ from the log-likelihood. This modification may also possess an impact on the behavior of the optimization algorithm, as e.g.~relative convergence criteria and step size could be affected.

\subsection{Simulation study}
\label{sec:simu_study}

The two previously analyzed data sets are both of comparably small size. In order to systematically investigate the performance of {\tt TMB} in the context of larger samples, we carried out a small simulation study. For this study, we simulated sequences of observations of length 2000 and 5000 from HMMs with two and three states, respectively. The parameters underlying the simulation are
\begin{equation*}
\bgamma =
\begin{pmatrix}
0.95 & 0.05\\
0.15 & 0.85
\end{pmatrix}, \quad
\bflambda = (1, 7),
\end{equation*}
for the two-state HMM and
\begin{equation*}
\bgamma =
\begin{pmatrix}
0.95 & 0.025 & 0.025\\
0.05 & 0.90 & 0.05\\
0.075 & 0.075 & 0.85
\end{pmatrix}, \quad
\bflambda = (1, 4, 7),
\end{equation*}
for the HMM with three states. \autoref{fig:data_plot_simu1} and \autoref{fig:data_plot_simu2} display the first 500 observations of two exemplary sequences of observations generated for the two- and three-state model, respectively.
{\hl We also simulated sequences of observations of length 2000 and 5000 from HMMs with four states. The parameters underlying both simulations are}
\begin{equation*}
\bgamma =
\begin{pmatrix}
0.85 & 0.05 & 0.05 & 0.05\\
0.05 & 0.85 & 0.05 & 0.05\\
0.05 & 0.10 & 0.80 & 0.05\\
0.034 & 0.033 & 0.033 & 0.90
\end{pmatrix}, \quad
\bflambda = (1, 5, 9, 13).
\end{equation*}

\begin{knitrout}
\definecolor{shadecolor}{rgb}{0.969, 0.969, 0.969}\color{fgcolor}\begin{figure}[htb]

{\centering \includegraphics[width=\maxwidth]{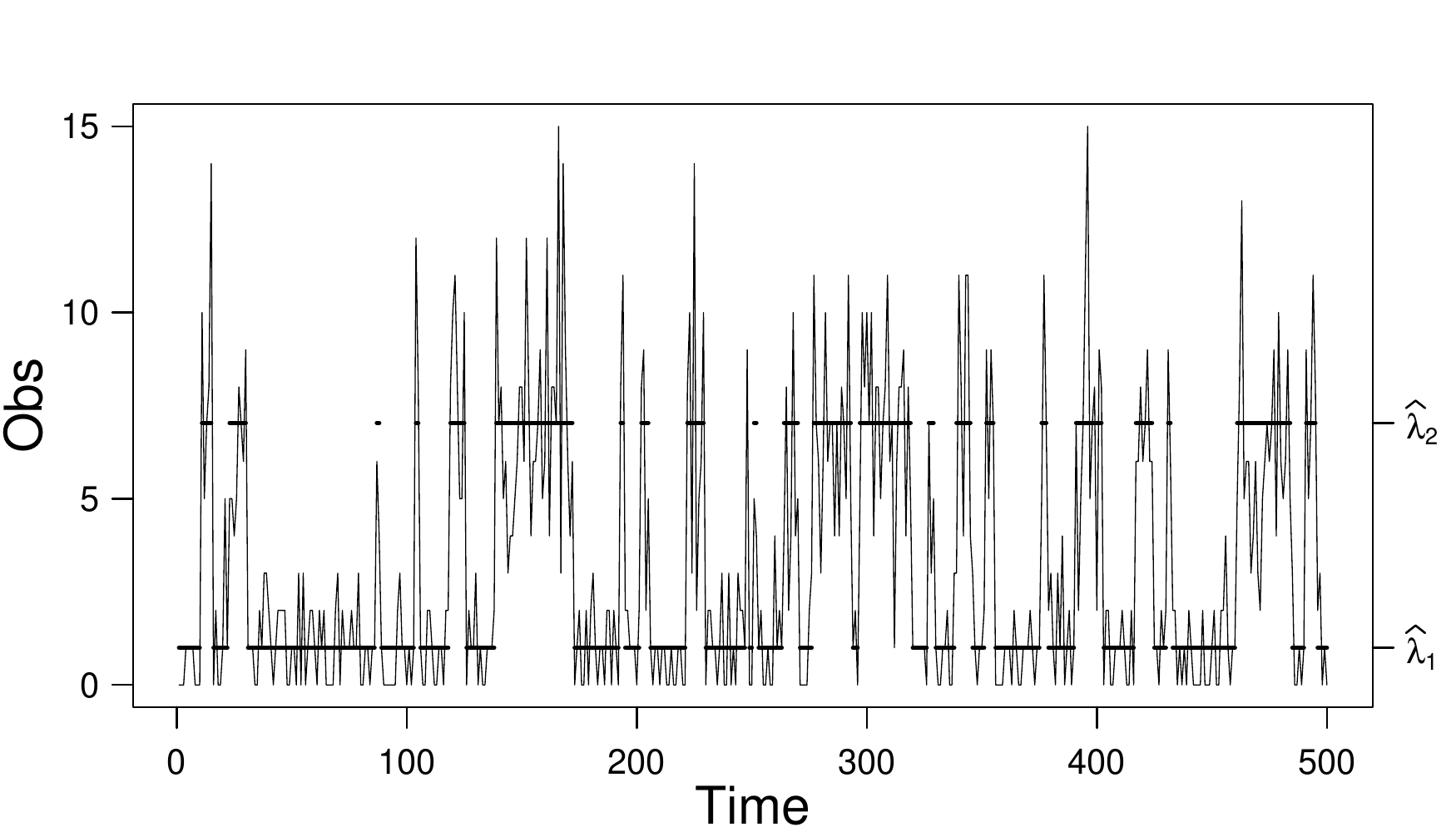} 

}

\caption{Plot of {\hl the first} simulated data (of size 2000) generated by a two-state Poisson HMM. The solid horizontal lines correspond to the conditional mean of the inferred state at each time. For readability, the graph is truncated to 500 data points. See \autoref{table:simu1_cis} for the values of $\widehat{\lambda}_i$.}\label{fig:data_plot_simu1}
\end{figure}

\end{knitrout}

\begin{knitrout}
\definecolor{shadecolor}{rgb}{0.969, 0.969, 0.969}\color{fgcolor}\begin{figure}[htb]

{\centering \includegraphics[width=\maxwidth]{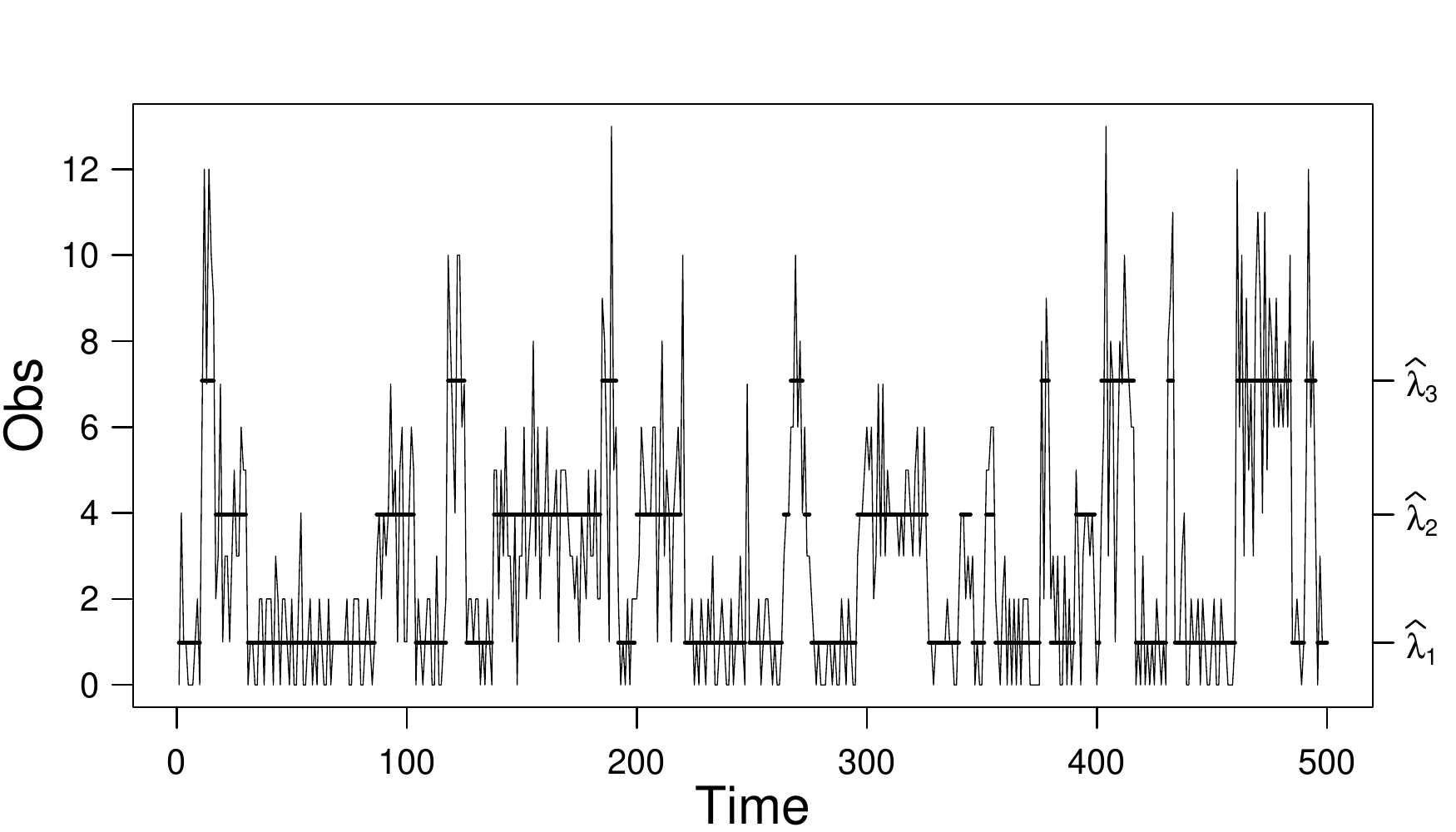} 

}

\caption{Plot of {\hl the second} simulated data (of size 5000) generated by a three-state Poisson HMM. The solid horizontal lines correspond to the conditional mean of the inferred state at each time. For readability, the graph is truncated to 500 data points. See \autoref{table:simu2_cis} for the values of $\widehat{\lambda}_i$.}\label{fig:data_plot_simu2}
\end{figure}

\end{knitrout}

\begin{knitrout}
\definecolor{shadecolor}{rgb}{0.969, 0.969, 0.969}\color{fgcolor}\begin{figure}[htb]

{\centering \includegraphics[width=\maxwidth]{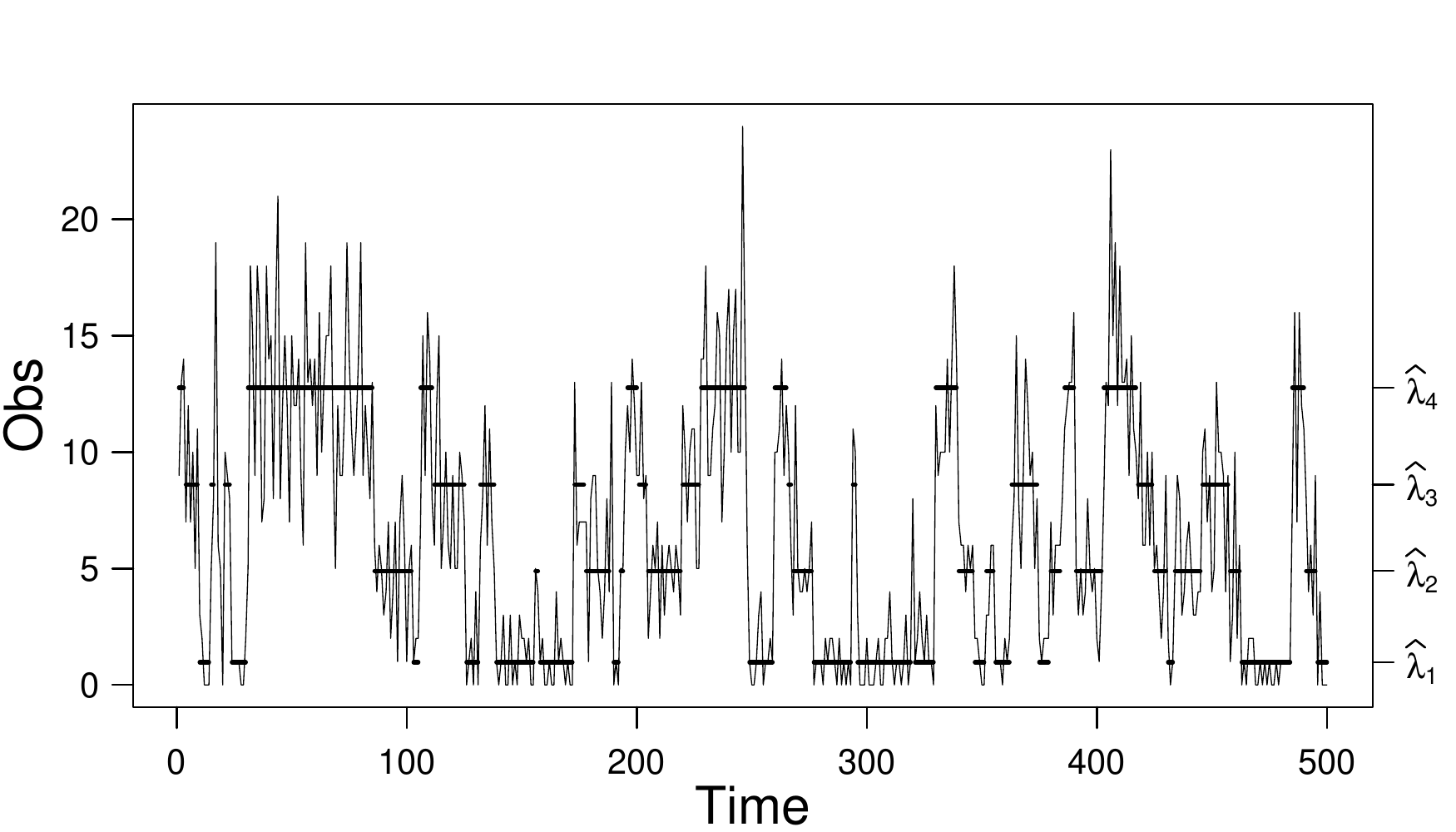} 

}

\caption{Plot of {\hl the third} simulated data (of size 2000) generated by a four-state Poisson HMM. The solid horizontal lines correspond to the conditional mean of the inferred state at each time. For readability, the graph is truncated to 500 data points. See \autoref{table:simu3_cis} for the values of $\widehat{\lambda}_i$.}\label{fig:data_plot_simu3}
\end{figure}

\end{knitrout}

\begin{knitrout}
\definecolor{shadecolor}{rgb}{0.969, 0.969, 0.969}\color{fgcolor}\begin{figure}[htb]

{\centering \includegraphics[width=\maxwidth]{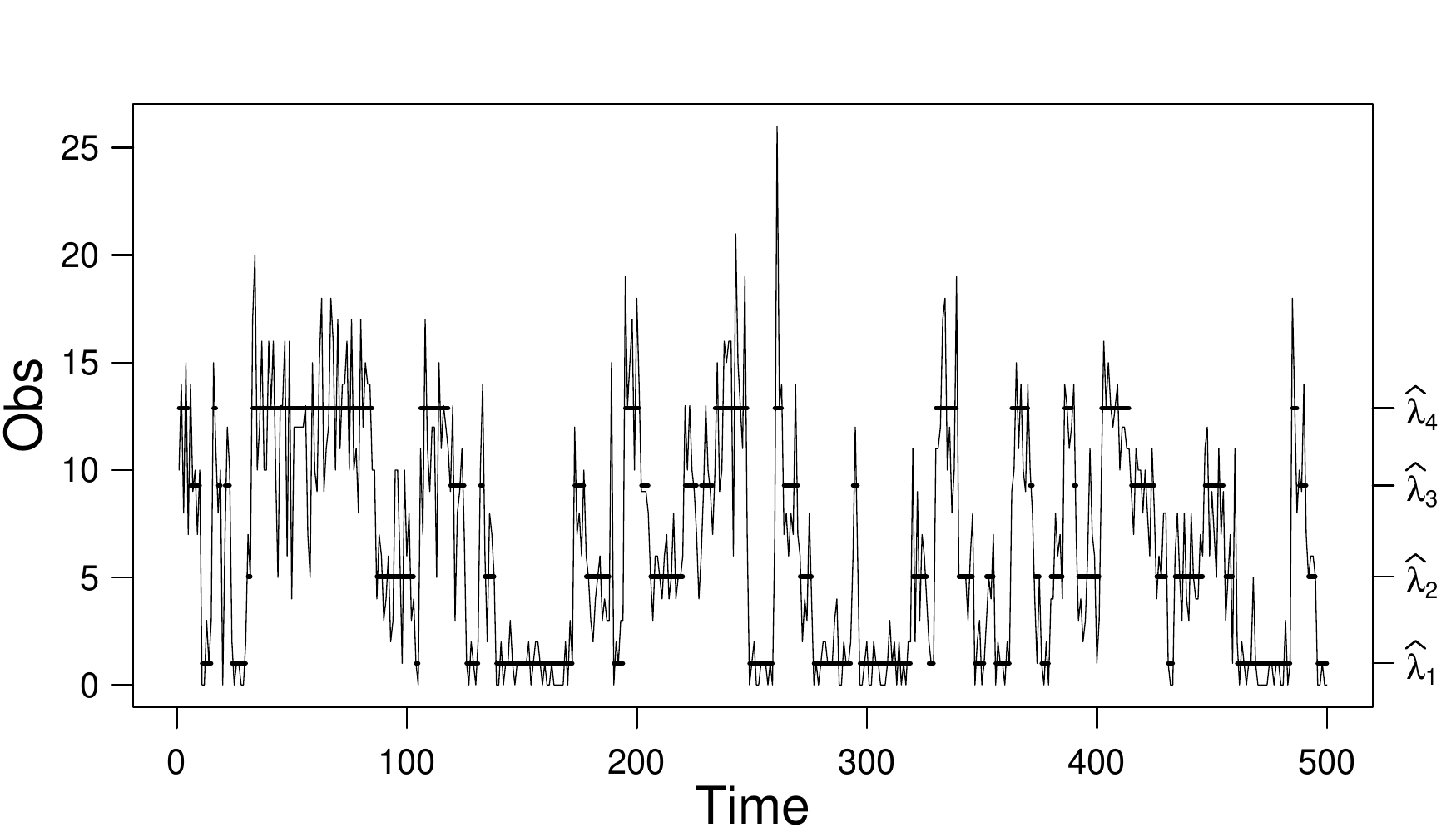} 

}

\caption{Plot of {\hl the fourth} simulated data (of size 5000) generated by a four-state Poisson HMM. The solid horizontal lines correspond to the conditional mean of the inferred state at each time. For readability, the graph is truncated to 500 data points. See \autoref{table:simu4_cis} for the values of $\widehat{\lambda}_i$.}\label{fig:data_plot_simu4}
\end{figure}

\end{knitrout}

The bootstrap setting described in the previous sections served for computing acceleration ratios for the simulated data sets. \autoref{table:speed-simu1} displays the results obtained for the two-state model. Again, $TMB_G$ provides the strongest acceleration compared to $DM$. Furthermore, $TMB_{GH}$ achieves the second place, closely followed by the remaining two approaches. This confirms the good performance of $TMB_G$ for large samples as well. Furthermore, it suggests a certain benefit from employing the Hessian computed by {\tt TMB} for longer observation sequences, since this method requires a much lower number of iterations than all other approaches.

\begin{table}[ht]
\centering
\begin{tabular}{lcccc}
  \hline
 & \textit{${TMB_0}$} & \textit{${TMB_G}$} & \textit{${TMB_H}$} & \textit{${TMB_{GH}}$} \\ 
  \hline
Acceleration ratio & 22.5 & 41.3 & 22.4 & 23.7 \\ 
   &  (20.6, 26.3)  &  (33, 50.6)  &  (20.6, 26.6)  &  (17.7, 31.2)  \\ 
  Iterations & 22.3 & 22.6 & 22.3 & 4.08 \\ 
   &  (17, 29)  &  (18, 28)  &  (17, 29)  &  (3, 6)  \\ 
   \hline
\end{tabular}
\caption{Acceleration and iterations for the {\hl first simulated data (of size 2000)}. The top lines show the acceleration ratios together with 95\% percentile bootstrap CIs when using {\tt TMB} in a bootstrap setting with 200 bootstrap samples. The bottom lines display the corresponding values for the number of iterations. {\hl The average parameter estimation duration without {\tt TMB} is 832 milliseconds.}} 
\label{table:speed-simu1}
\end{table}

The results for the three-state model, as shown in \autoref{table:speed-simu2}, basically underline all findings from the two-state model. Most importantly, again $TMB_{GH}$ requires a much lower number of iterations than the other approaches.

\begin{table}[ht]
\centering
\begin{tabular}{lcccc}
  \hline
 & \textit{${TMB_0}$} & \textit{${TMB_G}$} & \textit{${TMB_H}$} & \textit{${TMB_{GH}}$} \\ 
  \hline
Acceleration ratio & 14.2 & 54.1 & 14.2 & 31.4 \\ 
   &  (12.7, 15.7)  &  (44.4, 65.4)  &  (12.6, 15.9)  &  (24.4, 39.5)  \\ 
  Iterations & 42.4 & 42.6 & 42.4 & 4.28 \\ 
   &  (34, 50)  &  (34, 51.1)  &  (34, 50)  &  (3, 6)  \\ 
   \hline
\end{tabular}
\caption{Acceleration and iterations for the {\hl second simulated data (of size 5000)}. The top lines show the acceleration ratios together with 95\% percentile bootstrap CIs when using {\tt TMB} in a bootstrap setting with 200 bootstrap samples. The bottom lines display the corresponding values for the number of iterations. {\hl The average parameter estimation duration without {\tt TMB} is 8187 milliseconds.}} 
\label{table:speed-simu2}
\end{table}

\begin{table}[ht]
\centering
\begin{tabular}{lcccc}
  \hline
 & \textit{${TMB_0}$} & \textit{${TMB_G}$} & \textit{${TMB_H}$} & \textit{${TMB_{GH}}$} \\ 
  \hline
Acceleration ratio & 9.95 & 61.2 & 9.95 & 13.2 \\ 
   &  (9.38, 10.8)  &  (52.8, 69.5)  &  (9.37, 10.7)  &  (8.12, 18.6)  \\ 
  Iterations & 49.5 & 49.6 & 49.5 & 13 \\ 
   &  (38, 66)  &  (39, 67)  &  (38, 66)  &  (6, 24)  \\ 
   \hline
\end{tabular}
\caption{Acceleration and iterations for the {\hl third simulated data (of size 2000)}. The top lines show the acceleration ratios together with 95\% percentile bootstrap CIs when using {\tt TMB} in a bootstrap setting with 200 bootstrap samples. The bottom lines display the corresponding values for the number of iterations. {\hl The average parameter estimation duration without {\tt TMB} is 6777 milliseconds.}} 
\label{table:speed-simu3}
\end{table}

\begin{table}[ht]
\centering
\begin{tabular}{lcccc}
  \hline
 & \textit{${TMB_0}$} & \textit{${TMB_G}$} & \textit{${TMB_H}$} & \textit{${TMB_{GH}}$} \\ 
  \hline
Acceleration ratio & 12.2 & 74.3 & 12.2 & 29.1 \\ 
   &  (10.8, 13.7)  &  (62.9, 87.1)  &  (10.6, 13.6)  &  (21.5, 35.9)  \\ 
  Iterations & 51.7 & 51.8 & 51.7 & 5.62 \\ 
   &  (44, 58)  &  (44, 59)  &  (44, 58)  &  (4, 7.05)  \\ 
   \hline
\end{tabular}
\caption{Acceleration and iterations for the {\hl fourth simulated data (of size 5000)}. The top lines show the acceleration ratios together with 95\% percentile bootstrap CIs when using {\tt TMB} in a bootstrap setting with 200 bootstrap samples. The bottom lines display the corresponding values for the number of iterations. {\hl The average parameter estimation duration without {\tt TMB} is 21358 milliseconds.}} 
\label{table:speed-simu4}
\end{table}

Similar to the two previous sections, \autoref{table:simu1_cis} - \autoref{table:simu4_cis} show parameter estimates, CIs, and coverage probabilities.
The four exemplary sequences of observations shown in \autoref{fig:data_plot_simu1} - \autoref{fig:data_plot_simu4} served for deriving parameter estimates and CIs.
The coverage probabilities result from a Monte-Carlo study as the ones previously described.
For the two-state model, the CIs obtained by our three methods are very similar for all parameters.
The coverage probabilities lie comparably close to the theoretical level of 95\% for all parameters of the two-state model.
Moreover, no systematically too small or large CIs seem to result from any of the three methods.
The same holds true for the three-state model, with minor exceptions for the profile CIs.
For this method, the coverage probabilities of the CIs are close to 100\% for diagonal elements of the TPM, while the corresponding probabilities for off-diagonal elements are less than 95\%.
{\hl This pattern amplifies for the four-state models, as shown in \autoref{table:simu4_cis}.
Moreover, profile likelihood CIs of the elements of the TPM are not consistently provided by the \texttt{tmbprofile} function for the sequences of length 2000 from the four-state model. Therefore, the corresponding entries in \autoref{table:simu3_cis} remain empty.}

\begin{table}[ht]
\centering
\begin{tabular}{cccccccccccc}
  &&& \multicolumn{2}{c}{Wald-type CI}& \multicolumn{2}{c}{Profile CI}& \multicolumn{2}{c}{Bootstrap CI}& \multicolumn{3}{c}{Coverage prob. (\%)}\\ \hline
Par. & Value & Est. & L. & U. & L. & U. & L. & U. & {Wald} & Profile & Bootst. \\ 
  \hline
$\lambda_{1}$ & 1.00 & 0.99 & 0.93 & 1.04 & 0.93 & 1.04 & 0.93 & 1.04 & 95.1 & 95.4 & 94.9 \\ 
  $\lambda_{2}$ & 7.00 & 6.89 & 6.66 & 7.13 & 6.66 & 7.13 & 6.64 & 7.14 & 95.4 & 95.6 & 95.7 \\ 
  $\gamma_{11}$ & 0.95 & 0.94 & 0.93 & 0.95 & 0.92 & 0.95 & 0.92 & 0.95 & 95.1 & 95.1 & 95.1 \\ 
  $\gamma_{12}$ & 0.05 & 0.06 & 0.05 & 0.07 & 0.05 & 0.08 & 0.05 & 0.08 & 95.1 & 95.1 & 95.1 \\ 
  $\gamma_{21}$ & 0.15 & 0.16 & 0.13 & 0.20 & 0.13 & 0.20 & 0.13 & 0.20 & 94.1 & 95.1 & 94.5 \\ 
  $\gamma_{22}$ & 0.85 & 0.84 & 0.80 & 0.87 & 0.80 & 0.87 & 0.80 & 0.87 & 94.1 & 95.1 & 94.5 \\ 
  $\delta_{1}$ & 0.75 & 0.73 & 0.67 & 0.78 &  &  & 0.67 & 0.78 & 94.0 &  & 93.7 \\ 
  $\delta_{2}$ & 0.25 & 0.27 & 0.22 & 0.33 &  &  & 0.22 & 0.33 & 94.0 &  & 93.7 \\ 
   \hline
\end{tabular}
\caption{CIs for the {\hl first simulated dataset (of size 2000)}. From left to right, the columns contain: the parameter name, true parameter value, parameter estimate, and lower (L.) and upper (U.) bound of the corresponding 95\% CI derived via the Hessian provided by {\tt TMB}, likelihood profiling, and percentile bootstrap. Then follow coverage probabilities derived for these three methods in a Monte-Carlo study.} 
\label{table:simu1_cis}
\end{table}

\begin{table}[ht]
\centering
\begin{tabular}{cccccccccccc}
  &&& \multicolumn{2}{c}{Wald-type CI}& \multicolumn{2}{c}{Profile CI}& \multicolumn{2}{c}{Bootstrap CI}& \multicolumn{3}{c}{Coverage prob. (\%)}\\ \hline
Par. & Value & Est. & L. & U. & L. & U. & L. & U. & {Wald} & Profile & Bootst. \\ 
  \hline
$\lambda_{1}$ & 1.000 & 1.016 & 0.97 & 1.06 & 0.97 & 1.06 & 0.98 & 1.06 & 95.4 & 95.3 & 95.6 \\ 
  $\lambda_{2}$ & 4.000 & 4.014 & 3.86 & 4.17 & 3.86 & 4.17 & 3.86 & 4.16 & 93.6 & 93.7 & 94.2 \\ 
  $\lambda_{3}$ & 7.000 & 7.329 & 7.06 & 7.60 & 7.07 & 7.60 & 7.07 & 7.59 & 94.3 & 94.6 & 94.8 \\ 
  $\gamma_{11}$ & 0.950 & 0.950 & 0.94 & 0.96 & 0.93 & 0.97 & 0.94 & 0.96 & 95.3 & 100.0 & 95.1 \\ 
  $\gamma_{12}$ & 0.025 & 0.026 & 0.02 & 0.03 & 0.02 & 0.04 & 0.02 & 0.03 & 93.8 & 93.4 & 93.7 \\ 
  $\gamma_{13}$ & 0.025 & 0.024 & 0.02 & 0.03 & 0.02 & 0.03 & 0.02 & 0.03 & 94.5 & 94.2 & 95.0 \\ 
  $\gamma_{21}$ & 0.050 & 0.050 & 0.03 & 0.06 & 0.04 & 0.06 & 0.04 & 0.07 & 95.2 & 93.1 & 95.2 \\ 
  $\gamma_{22}$ & 0.900 & 0.906 & 0.89 & 0.93 & 0.87 & 0.93 & 0.88 & 0.93 & 95.2 & 99.5 & 95.0 \\ 
  $\gamma_{23}$ & 0.050 & 0.045 & 0.03 & 0.06 & 0.03 & 0.06 & 0.03 & 0.06 & 94.1 & 93.3 & 94.8 \\ 
  $\gamma_{31}$ & 0.075 & 0.077 & 0.05 & 0.10 & 0.06 & 0.10 & 0.05 & 0.10 & 95.2 & 92.1 & 95.9 \\ 
  $\gamma_{32}$ & 0.075 & 0.089 & 0.06 & 0.12 & 0.06 & 0.12 & 0.06 & 0.13 & 94.9 & 93.5 & 95.2 \\ 
  $\gamma_{33}$ & 0.850 & 0.834 & 0.80 & 0.87 & 0.78 & 0.88 & 0.79 & 0.87 & 95.0 & 99.6 & 95.1 \\ 
  $\delta_{1}$ & 0.545 & 0.541 & 0.48 & 0.60 &  &  & 0.49 & 0.60 & 94.2 &  & 94.2 \\ 
  $\delta_{2}$ & 0.273 & 0.300 & 0.25 & 0.35 &  &  & 0.25 & 0.35 & 94.7 &  & 94.7 \\ 
  $\delta_{3}$ & 0.182 & 0.159 & 0.12 & 0.19 &  &  & 0.13 & 0.20 & 93.9 &  & 94.4 \\ 
   \hline
\end{tabular}
\caption{CIs for the {\hl second simulated dataset (of size 5000)}. From left to right, the columns contain: the parameter name, true parameter value, parameter estimate, and lower (L.) and upper (U.) bound of the corresponding 95\% CI derived via the Hessian provided by {\tt TMB}, likelihood profiling, and percentile bootstrap. Then follow coverage probabilities derived for these three methods in a Monte-Carlo study.} 
\label{table:simu2_cis}
\end{table}

\begin{table}[ht]
\centering
\begin{tabular}{cccccccccccc}
  &&& \multicolumn{2}{c}{Wald-type CI}& \multicolumn{2}{c}{Profile CI}& \multicolumn{2}{c}{Bootstrap CI}& \multicolumn{3}{c}{Coverage prob. (\%)}\\ \hline
Par. & Value & Est. & L. & U. & L. & U. & L. & U. & {Wald} & Profile & Bootst. \\ 
  \hline
$\lambda_{1}$ & 1.000 & 1.024 & 0.90 & 1.15 &  &  & 0.90 & 1.16 & 95.7 & 95.7 & 95.9 \\ 
  $\lambda_{2}$ & 5.000 & 5.298 & 5.01 & 5.59 &  &  & 4.99 & 5.61 & 93.7 & 93.9 & 94.4 \\ 
  $\lambda_{3}$ & 9.000 & 10.359 & 9.61 & 11.11 &  &  & 9.52 & 11.13 & 92.4 & 93.7 & 95.9 \\ 
  $\lambda_{4}$ & 13.000 & 13.451 & 12.98 & 13.92 &  &  & 13.01 & 14.00 & 95.1 & 95.3 & 96.1 \\ 
  $\gamma_{11}$ & 0.850 & 0.815 & 0.77 & 0.86 &  &  & 0.76 & 0.86 & 94.8 &  & 94.0 \\ 
  $\gamma_{12}$ & 0.050 & 0.061 & 0.03 & 0.10 &  &  & 0.03 & 0.10 & 93.8 &  & 94.7 \\ 
  $\gamma_{13}$ & 0.050 & 0.026 & 0.00 & 0.07 &  &  & 0.00 & 0.07 & 93.7 &  & 94.6 \\ 
  $\gamma_{14}$ & 0.050 & 0.097 & 0.05 & 0.14 &  &  & 0.05 & 0.15 & 93.2 &  & 94.9 \\ 
  $\gamma_{21}$ & 0.050 & 0.062 & 0.04 & 0.09 &  &  & 0.04 & 0.09 & 92.7 &  & 94.0 \\ 
  $\gamma_{22}$ & 0.850 & 0.847 & 0.81 & 0.89 &  &  & 0.80 & 0.88 & 95.4 &  & 95.4 \\ 
  $\gamma_{23}$ & 0.050 & 0.061 & 0.02 & 0.10 &  &  & 0.02 & 0.11 & 90.0 &  & 94.3 \\ 
  $\gamma_{24}$ & 0.050 & 0.031 & 0.00 & 0.06 &  &  & 0.00 & 0.07 & 92.7 &  & 94.1 \\ 
  $\gamma_{31}$ & 0.050 & 0.033 & 0.00 & 0.06 &  &  & 0.00 & 0.07 & 91.9 &  & 93.0 \\ 
  $\gamma_{32}$ & 0.100 & 0.107 & 0.05 & 0.16 &  &  & 0.06 & 0.17 & 93.9 &  & 94.7 \\ 
  $\gamma_{33}$ & 0.800 & 0.827 & 0.76 & 0.90 &  &  & 0.74 & 0.88 & 94.3 &  & 96.0 \\ 
  $\gamma_{34}$ & 0.050 & 0.033 & 0.00 & 0.07 &  &  & 0.00 & 0.09 & 89.5 &  & 95.4 \\ 
  $\gamma_{41}$ & 0.034 & 0.022 & 0.00 & 0.04 &  &  & 0.00 & 0.04 & 93.8 &  & 94.1 \\ 
  $\gamma_{42}$ & 0.033 & 0.023 & 0.00 & 0.05 &  &  & 0.00 & 0.05 & 91.7 &  & 93.4 \\ 
  $\gamma_{43}$ & 0.033 & 0.058 & 0.01 & 0.10 &  &  & 0.02 & 0.12 & 89.4 &  & 93.5 \\ 
  $\gamma_{44}$ & 0.900 & 0.896 & 0.86 & 0.93 &  &  & 0.85 & 0.93 & 95.1 &  & 94.4 \\ 
  $\delta_{1}$ & 0.223 & 0.173 & 0.13 & 0.22 &  &  & 0.13 & 0.22 & 93.1 &  & 93.2 \\ 
  $\delta_{2}$ & 0.266 & 0.278 & 0.22 & 0.34 &  &  & 0.21 & 0.34 & 94.1 &  & 94.8 \\ 
  $\delta_{3}$ & 0.177 & 0.231 & 0.15 & 0.31 &  &  & 0.15 & 0.32 & 93.8 &  & 96.9 \\ 
  $\delta_{4}$ & 0.333 & 0.318 & 0.22 & 0.42 &  &  & 0.21 & 0.42 & 93.6 &  & 93.8 \\ 
   \hline
\end{tabular}
\caption{CIs for the {\hl third simulated dataset (of size 2000)}. From left to right, the columns contain: the number of hidden states, parameter name, true parameter value, parameter estimate, and lower (L.) and upper (U.) bound of the corresponding 95\% CI derived via the Hessian provided by {\tt TMB}, likelihood profiling, and percentile bootstrap. Then follow coverage probabilities derived for these three methods in a Monte-Carlo study.} 
\label{table:simu3_cis}
\end{table}

\begin{table}[ht]
\centering
\begin{tabular}{cccccccccccc}
  &&& \multicolumn{2}{c}{Wald-type CI}& \multicolumn{2}{c}{Profile CI}& \multicolumn{2}{c}{Bootstrap CI}& \multicolumn{3}{c}{Coverage prob. (\%)}\\ \hline
Par. & Value & Est. & L. & U. & L. & U. & L. & U. & {Wald} & Profile & Bootst. \\ 
  \hline
$\lambda_{1}$ & 1.000 & 0.955 & 0.89 & 1.02 & 0.89 & 1.02 & 0.89 & 1.03 & 96.0 & 95.9 & 95.7 \\ 
  $\lambda_{2}$ & 5.000 & 5.192 & 5.00 & 5.39 & 4.99 & 5.38 & 4.99 & 5.40 & 95.2 & 95.1 & 95.2 \\ 
  $\lambda_{3}$ & 9.000 & 9.253 & 8.65 & 9.86 & 8.66 & 9.86 & 8.63 & 9.85 & 93.4 & 93.6 & 94.6 \\ 
  $\lambda_{4}$ & 13.000 & 12.987 & 12.74 & 13.24 & 12.75 & 13.25 & 12.74 & 13.27 & 95.5 & 95.3 & 95.1 \\ 
  $\gamma_{11}$ & 0.850 & 0.840 & 0.82 & 0.86 & 0.78 & 0.89 & 0.82 & 0.86 & 95.1 & 99.9 & 95.0 \\ 
  $\gamma_{12}$ & 0.050 & 0.064 & 0.04 & 0.08 & 0.05 & 0.08 & 0.04 & 0.09 & 94.4 & 92.5 & 95.2 \\ 
  $\gamma_{13}$ & 0.050 & 0.044 & 0.02 & 0.07 & 0.02 & 0.06 & 0.02 & 0.07 & 93.4 & 90.4 & 93.8 \\ 
  $\gamma_{14}$ & 0.050 & 0.052 & 0.03 & 0.07 & 0.04 & 0.07 & 0.03 & 0.07 & 94.5 & 89.6 & 94.4 \\ 
  $\gamma_{21}$ & 0.050 & 0.058 & 0.04 & 0.08 & 0.04 & 0.07 & 0.04 & 0.08 & 95.0 & 91.2 & 95.3 \\ 
  $\gamma_{22}$ & 0.850 & 0.836 & 0.81 & 0.86 & 0.78 & 0.89 & 0.80 & 0.86 & 94.8 & 100.0 & 95.2 \\ 
  $\gamma_{23}$ & 0.050 & 0.049 & 0.02 & 0.08 & 0.03 & 0.08 & 0.02 & 0.08 & 94.7 & 92.2 & 95.1 \\ 
  $\gamma_{24}$ & 0.050 & 0.057 & 0.04 & 0.08 & 0.04 & 0.08 & 0.04 & 0.08 & 95.0 & 91.8 & 95.8 \\ 
  $\gamma_{31}$ & 0.050 & 0.057 & 0.02 & 0.09 & 0.03 & 0.08 & 0.03 & 0.09 & 96.1 & 90.0 & 95.9 \\ 
  $\gamma_{32}$ & 0.100 & 0.098 & 0.05 & 0.15 & 0.06 & 0.14 & 0.05 & 0.15 & 94.3 & 90.3 & 94.5 \\ 
  $\gamma_{33}$ & 0.800 & 0.786 & 0.73 & 0.84 & 0.68 & 0.89 & 0.71 & 0.84 & 94.4 & 100.0 & 95.2 \\ 
  $\gamma_{34}$ & 0.050 & 0.059 & 0.02 & 0.10 & 0.02 & 0.10 & 0.01 & 0.12 & 97.0 & 94.2 & 96.7 \\ 
  $\gamma_{41}$ & 0.034 & 0.029 & 0.02 & 0.04 & 0.02 & 0.04 & 0.02 & 0.04 & 94.5 & 91.2 & 94.3 \\ 
  $\gamma_{42}$ & 0.033 & 0.050 & 0.03 & 0.07 & 0.03 & 0.07 & 0.03 & 0.07 & 95.8 & 93.2 & 96.1 \\ 
  $\gamma_{43}$ & 0.033 & 0.028 & 0.01 & 0.05 & 0.01 & 0.05 & 0.01 & 0.06 & 95.0 & 94.7 & 96.2 \\ 
  $\gamma_{44}$ & 0.900 & 0.892 & 0.87 & 0.91 & 0.84 & 0.94 & 0.87 & 0.91 & 95.7 & 100.0 & 95.6 \\ 
  $\delta_{1}$ & 0.223 & 0.220 & 0.19 & 0.25 &  &  & 0.19 & 0.25 & 94.7 &  & 94.8 \\ 
  $\delta_{2}$ & 0.266 & 0.283 & 0.24 & 0.32 &  &  & 0.24 & 0.32 & 94.4 &  & 94.8 \\ 
  $\delta_{3}$ & 0.177 & 0.154 & 0.11 & 0.19 &  &  & 0.11 & 0.20 & 96.6 &  & 96.7 \\ 
  $\delta_{4}$ & 0.333 & 0.342 & 0.29 & 0.40 &  &  & 0.29 & 0.39 & 94.1 &  & 94.1 \\ 
   \hline
\end{tabular}
\caption{CIs for the {\hl fourth simulated dataset (of size 5000)}. From left to right, the columns contain: the parameter name, true parameter value, parameter estimate, and lower (L.) and upper (U.) bound of the corresponding 95\% CI derived via the Hessian provided by {\tt TMB}, likelihood profiling, and percentile bootstrap. Then follow coverage probabilities derived for these three methods in a Monte-Carlo study.} 
\label{table:simu4_cis}
\end{table}

{\hl Last, we would like to point out that the runtime of our entire simulation study was approximately one week.
Without the acceleration by {\tt{TMB}}, several months would have been necessary.
This illustrates not only the saving of time but also of resources (e.g. energy consumption, wear of IT-infrastructure).}

\clearpage
\section{Discussion}
\label{sec:discussion}

In this tutorial, we provide researchers from all applied fields with an introduction to parameter estimation for HMMs via {\tt{TMB}} using {\tt{R}}.
Some procedures need to be coded in {\tt{C++}}, {\hl which represents a certain requirement on the user and may not be beneficial if, e.g., an HMM needs to be estimated only once. However, for users working more or less regularly with HMMs,} the use of {\tt{TMB}} accelerates existing parameter estimation procedures without having to carry out major changes to {\tt{R}} code that is already in use. Moreover, after finishing the estimation procedure, {\tt{TMB}} obtains standard errors for the estimated parameters at a very low computational cost.

We examined the performance of {\tt{TMB}} in the context of Poisson HMMs through two small real data sets and in a simulation setting with longer sequences of observations. Overall, it is notable that the parameter estimation process is strongly kcelerated on the one hand. This applies even to small data sets, and the highest acceleration is obtained when only the gradient is supplied by {\tt{TMB}} (instead of both gradient and Hessian). On the other hand, the standard errors obtained through {\tt{TMB}} are very similar to the standard errors obtained by profiling the likelihood and bootstrapping while being (much) less computationally intensive. This is novel since Hessian-based CIs did not seem to be reliable in the past, as illustrated e.g.~by \cite{visser}.

Along with the tutorial character of this paper comes the shortcoming that we restricted ourselves to only one comparably simple HMM with Poisson conditional distributions. The extension to other distributions is, however, not overly complicated. We briefly illustrate the case of Gaussian conditional distributions on the supporting GitHub page. Moreover, to keep the paper at acceptable length, we were not able to address a couple of potential extensions and research questions.
{\hl For example, it would be interesting to investigate whether supplying derivatives provided by {\tt{TMB}} to different optimizers has a positive impact on convergence properties.
This may be of particular interest when considering the impact of different potentially poor initial values on the optimization results.
Following} \citet{bulla}{\hl, one could also consider a hybrid algorithm in this context to benefit from the advantageous properties of the EM algorithm, such as the resilience to poor initial values.
When setting up this approach, the EM algorithm would most likely also benefit from acceleration by C++.}
The reliability of Wald-type CIs provided by {\tt{TMB}} for other models and very long sequences with e.g.~hundreds of thousands or millions of observations could also be of interest.
{\hl In addition, improving the reliability of CIs for small samples (such as our TYT and lamb data) may be worth investigating.}
In a similar direction, it seems rather obvious to benefit from {\tt{TMB}} for more complex settings such as panel data with random effects, where computationally efficient routines play an important role.
Furthermore, in the bootstrap analysis of the TYT and in particular of the lamb data set, we noted that the {\tt{TMB}}-internal function {\tt tmbprofile} sometimes fails to provide CIs for parameters very close to a boundary. For these two data sets, this was the case for the elements of the TPM which take values close to one: approximately 7\% (TYT) and 29\% (lamb), respectively, of the generated bootstrap samples were affected.
{\hl The same problem showed for the four-state model with 2000 observations (here more than 50\% of the generated data sets).}
It remains to clarify whether this problem can be solved by a suitable modification of \texttt{tmbprofile}, or if the underlying difficulties require an entirely different approach.
{\hl
Last, profile likelihood CIs for the elements of the TPM seems to be subject to bias for models with more than two states.
This may be due to the strong dependence of all elements of the TPM in the same row, which is problematic for a proper definition of the profile likelihood function $L_p(\gamma_{ij})$ difficult \citep[see, e.g.,][]{fischer}.
Therefore, we recommend treating profile CIs for the elements of the TPM with care, in particular for models with more than two states, and further research in this direction is needed.
}

From an application perspective, the use of {\tt{TMB}} allows executing estimation procedures at a significantly reduced cost compared to the execution of plain {\tt{R}}. Such a performance gain could be of interest when repeatedly executing statistical procedures on mobile devices. It seems plausible to enrich the TYT app (or similar apps collecting a sufficient amount of data) by integrating a warning system that detects when the user enters a new state, inferred through a suitable HMM or another procedure accelerated via {\tt{TMB}} in real-time. This new state could, e.g., represent an improvement or worsening of a pre-defined medical condition and recommend the user to contact the consulting physician if the change persists for a certain period. Provided the agreement of the user, this collected information could also be pre-processed and transferred automatically to the treating physician and allow to identify personalized treatment options.

\section*{Acknowledgement}
The work of J.~Bulla was supported by the GENDER-Net Co-Plus Fund (GNP-182). Furthermore, this work was supported by the Financial Market Fund (Norwegian Research Council project no.~309218). We thank the University of Regensburg and European School for Interdisciplinary Tinnitus Research (ESIT) for providing access to the TYT data. Special thanks go to J.~Sim\~{o}es for data preparation, and W.~Schlee and B.~Langguth for helpful comments and suggestions. J.~Bulla's profound thanks for moral support go to the highly skilled, very helpful, and always friendly employees Vestbrygg AS (Org.~ID 912105954).

\vspace*{1pc}

\noindent {\bf{Conflict of Interest}}

\noindent {\it{The authors have declared no conflict of interest.}}

\clearpage
\bibliographystyle{apalike}
\bibliography{paper,packages}

\end{document}